\documentclass[aps,pre,reprint,a4paper,showpacs]{revtex4-1}

\usepackage{amsmath}
\usepackage{amssymb}

\usepackage{graphicx}

\usepackage{color}

\newcommand{\MSD}{$\left\langle R^{2}(t) \right\rangle$ }
\newcommand{\MSDt}{$\left\langle R^{2}(t) \right\rangle/t$ }

\begin{document}

\title{Anomalous diffusion due to hindering by mobile obstacles undergoing Brownian motion or Orstein-Ulhenbeck processes.}
\date{\today}

\author{Hugues Berry}
\email[Send correspondence to: ]{hugues.berry@inria.fr}
\affiliation{EPI Beagle, INRIA Rh\^one-Alpes, F-69603, Villeurbanne, France}
\affiliation{LIRIS, Universit\'e de Lyon, UMR 5205 CNRS-INSA, F-69621, Villeurbanne, France}
\author{Hugues Chat\'{e}}
\affiliation{Service de Physique de l'Etat Condens\'{e}, URA2464, CEA-Saclay, 91191 Gif-sur-Yvette, France}

\begin{abstract}
In vivo measurements of the passive movements of biomolecules or vesicles in cells consistently report ``anomalous diffusion'', where mean-squared displacements scale as a power law of time with exponent $\alpha< 1$ (subdiffusion). While the detailed mechanisms causing such behaviors are not always elucidated, movement hindrance by obstacles is often invoked. However, our understanding of how hindered diffusion leads to subdiffusion is based on diffusion amidst randomly-located \textit{immobile} obstacles. Here, we have used Monte-Carlo simulations to investigate transient subdiffusion due to \textit{mobile} obstacles with various modes of mobility. Our simulations confirm that the anomalous regimes rapidly disappear when the obstacles move by Brownian motion. By contrast, mobile obstacles with more confined displacements, e.g. Orstein-Ulhenbeck motion, are shown to preserve subdiffusive regimes. The mean-squared displacement of tracked protein displays convincing power-laws with anomalous exponent $\alpha$ that varies with the density of OU obstacles or the relaxation time-scale of the OU process. In particular, some of the values we observed are significantly below the universal value predicted for immobile obstacles in 2d. Therefore, our results show that subdiffusion due to mobile obstacles with OU-type of motion may account for the large variation range exhibited by experimental measurements in living cells and may explain that some experimental estimates are below the universal value predicted for immobile obstacles.
\end{abstract}

\maketitle
\section{\label{sec:intro}INTRODUCTION}
The inner life of a cell involves complex reaction-diffusion processes whereby biomolecules interact with each other. Because biomolecules interact only when they meet, the way by which they actually move, i.e. the diffusion part of these processes, has a deep impact. More often than not, it is hypothesised that the intracellular micro-environment is very simple, so that biomolecule movement can be described by classical Brownian motion, a hallmark of which is the linear relation between the average 
of the squared distance travelled by the molecule 
(mean squared displacement) and time: 
$\left\langle R^{2}(t)\right\rangle\propto t$. 
By contrast, experimental and vesicular measurements of molecular diffusion in living cells have consistently reported nonlinear relations
in almost all cell compartments, either in procaryotes or eucaryote \cite{Schwille1999,Smith1999,Wachsmuth2000,Seisenberger2001,Fujiwara2002,Caspi2002,Platani2002, Tolic-Noerrelykke2004,Golding2006,Bronstein2009,Weber2010,Jeon2011,Weigel2011,Tabei2013} (see H\"{o}fling and Franosch, 2013 \cite{Hoefling2013} for a recent review). 
Most often, these nonlinear variations are found to be power laws 
($\left\langle R^{2}(t)\right\rangle\propto t^\alpha$ with $\alpha \neq 1$). Super-diffusive motion ($2>\alpha > 1$) is relatively well 
understood, being usually due to active transport mediated by molecular 
motors on cytoskeleton elements~\cite{Kulic2008},
but subdiffusive transport ($\alpha < 1$) 
less so. 

In bacterial cytoplasm, small macromolecules, ranging from small proteins like GFP to intermediate-sized protein aggregates, seem to display Brownian motion ($\alpha=1$) \cite{Elowitz1999,*Bakshi2011,English2011,Coquel2013}, but the motion of larger biomolecules, such as RNA particles or ribosomes is subdiffusive \cite{Golding2006,Weber2010,English2011}. In the cytoplasm of mammal cells, the motion is subdiffusive for a large range of sizes, from large objects (beads, dextrans, granules) \cite{Weiss2004,Guigas2007,Guigas2007a,Jeon2011,Tabei2013} down to small proteins \cite{Wachsmuth2000}. The reported values of $\alpha$ vary over a wide interval (between 0.5 and 0.9), even for molecules of similar size. In the plasma membrane of mammal cells, the reported values also consistently exhibit subdiffusion, with a similar variation range for the exponent \cite{Schwille1999,Murase2004}, in particular when receptor motion \cite{Feder1996,Smith1999,Vrljic2002,Weigel2011} is considered (between 0.49 and 0.9). Subdiffusion has also been reported in the nucleus for a large range of diffusive object size, from small proteins (GFP and fusion thereof) \cite{Wachsmuth2000} to large complexes (Cajal bodies, telomeres) \cite{Platani2002,Guigas2007a,Bronstein2009}. In this case as well, the estimated values of $\alpha$ take values within a very large variation range (between 0.32 and 0.9). 

There exist three major theoretical scenarios to explain subdiffusive transport, all of which rest on the idea that the interior of cells and their membranes experience large molecular crowding due to their high densities of proteins, lipids, carbohydrates, filamentous networks and organelles, with widely-distributed sizes~\cite{Dix2008}. In the presence of a hierarchy of such slow processes slowing-down diffusion, one generically expects subdiffusion. Compared to experimental data evidencing subdiffusion, it can be nontrivial to decide which of these three scenarios matches the data~\cite{Condamin2008}, especially because these three scenario need not to be mutually exclusive and must sometimes be combined to account for the experimental observations \cite{Weigel2011,Tabei2013}. 

The arguably simplest scenario, referred to as ``Fractional Brownian Motion'', is a generalization of the classical Brownian motion, where the random increments between two successive locations are not independent (like in Brownian motion) but present long-range temporal correlations~\cite{barkai-phystoday-2012}. The second scenario, usually referred to as ``Continuous-Time Random Walks'' (CTRW) assumes that the complexity of the cellular media changes the statistics of the residence time between two moves of the random walkers. Whereas Dirac- or exponentially-distributed residence times lead to the classical Brownian motion, power-law distributed residence times can generate non-equilibrium processes with subdiffusive motion~\cite{Bouchaud1990,Metzler2000}. 

The third scenario is the only one to provide a clear microscopic origin to the observed subdiffusion. It assumes that intracellular movements are restricted (by e.g. molecular crowding) to a subset of the cellular space that has fractal geometry. Random walks restricted to fractal supports are indeed known to exhibit subdiffusion~\cite{Bouchaud1990,Renner2005}. One acclaimed model for this scenario is hindered diffusion in the presence of randomly-distributed immobile obstacles~\cite{Saxton1994,Berry2002,Kammerer2008,Hoefling2008a,Bauer2010,Hoefling2013}. When obstacle density is at the percolation threshold, the subspace available to the diffusing molecule forms a percolation cluster and the mean-square displacement scales sublinearly with time as $\left\langle R^{2}(t)\right\rangle\propto t^\alpha$ with (in continuum space) $\alpha = 0.659$ in 2d and $ 0.317$ in 3d~\cite{Bouchaud1990,Kammerer2008}. When obstacle density is lower, subdiffusion is only transient: at long time scales, diffusion leaves the  subdiffusion regime and converges  to slowed-down Brownian motion. But the value of $\alpha$ during the anomalous regime is not expected to change~\cite{Bouchaud1990,Kammerer2008,Soula2013}. 

Although hinderance by immobile obstacles is a seductive scenario to subdiffusion, this scenario predicts that the anomalous exponent $\alpha$ has a universal, thus unique value that varies only when the dimensionality of the problem or the discrete vs continuum properties of space (in 3d) changes \cite{Bouchaud1990,Kammerer2008}. The above reported large variation range of the experimental measurements of $\alpha$ in cells is therefore hard to reconcile with this scenario. In particular, some of the reported values are significantly \textit{smaller} than the theoretical values: in cell membranes, estimates of $\alpha \approx 0.5$ have been reported \cite{Smith1999,Murase2004}, a value significantly smaller than the universal value of 0.659 predicted by the immobile obstacle scenario. 

The assumption that obstacles are immobile is very practical both because it makes simulation much easier and efficient and because it permits direct application of percolation theory. Albeit on general grounds, obstacles can be expected to be less mobile than the tracked molecules because of their size, assuming their total immobility is a strong hypothesis that deserves further investigation. Some studies have been devoted to the case where the obstacles undergo Brownian motion and concluded that even when obstacle motion is much slower than the tracked molecule, the transient subdiffusion regime should rapidly vanish~\cite{Saxton1994,Tremmel2003}. Brownian motion is however not the only motion possible in cells, as evidenced by the experimental results reported above, and other types of obstacle motion must be considered.

In the present work, we investigate subdiffusion due to mobile obstacles in two space dimensions. Using Monte-Carlo simulations, we show that the effect of obstacle movements depends on the type of movement considered. While obstacles endowed with Brownian motion efficiently suppress the subdiffusive regime, it is preserved when obstacle movement is more confined than Brownian motion. To emulate this confinement, we use an Ornstein-Uhlenbeck (OU) process (a Brownian motion coupled to a slow drift to the long-term mean position) to model obstacle motion. Our results show that when obstacle motion is described by an OU process, the subdiffusive regime is conserved, even for for large obstacle mobilities. Moreover, when the density of OU obstacles varies above the percolation threshold for immobile obstacles, our simulations show convincing evidence of subdiffusion regimes with values of $\alpha$ that depend on obstacle density and are within the experimental range. Therefore, our results show that accounting for obstacle motion by OU-processes qualifies hindered diffusion as a potential microscopic mechanism for the experimental observations of subdiffusion in cells.

\section{\label{sec:Methods}METHODS}

\subsection{Diffusion constants and molecule radii}
The typical values of the protein and obstacle sizes and mobility in our simulations were chosen so as to be representative of the size encountered in a typical cell. 
Regarding the size of the diffusing protein, we considered an ``average'' {\it E. coli} protein, that typically has 
radius $r_{\rm w}=2.0 \mathrm{\;n m}$ and molecular weight $40\mathrm{\;kDa}$, see e.g. Table S1 in~\cite{McGuffee2010}. Typical lateral diffusion coefficients for such ``average-sized'' bacterial proteins range from $10^{2}\mathrm{\;\mu m^2/s}$ in (unobstructed) water to $10^{0}-10^{1}\mathrm{\;\mu m^2/s}$ in the (obstructed) {\it E coli} cytoplasm~\cite{Elowitz1999,Nenninger2010}. In two dimensions however, typical orders of magnitude vary from $10^{0}\mathrm{\;\mu m^2/s}$ in (unobstructed) artificial membranes down to $10^{-2}-10^{-1}\mathrm{\;\mu m^2/s}$ in (obstructed) cytoplasmic membranes~\cite{Jacobson1987}. Since our simulations are two-dimensional we focused on the later case. The diffusion constant in our simulations corresponds to diffusion in the membrane without obstacle, therefore we set the protein diffusion constant to $D_{\rm RW}=1.0 \mathrm{\;\mu m^2/s}$. Regarding the obstacles, we considered large multimolecular obstacles (comparable to ribosomes), with radius $r_{\rm obs}=5 \mathrm{\;n m}$.

\subsection{A continuum percolation model}
We simulated the two-dimensional diffusion of proteins in lattice-free conditions. Periodic boundary conditions were used to reduce finite-size effects. 
Each run was initiated by positioning
at random (with uniform probability) 
2d obstacles (disks) in the 2d continuous space domain of overall size 
$L_{x}\times L_{y}=5.0\times 5.0 \;\mathrm{\mu m^{2}}$, until the surface fraction occupied by the 
obstacles equals the preset excluded volume fraction $\theta$. A set of 
$N_{\mathrm{w}}=10$ non-interacting proteins (random walkers) was then positioned in the space 
domain at random locations but respecting excluded volume w.r.t. obstacles: 
for each protein, a new position $\mathbf{x}_{\rm w}$ is chosen at random 
(with uniform distribution) inside the simulation domain, until the tracer molecule does not overlap with any obstacle at  $\mathbf{x}_{\rm w}$. \\
Excluded volume was thus imposed between proteins and obstacles (i.e. one protein and one obstacle cannot share the same spatial region) but not between two proteins nor between two obstacles. This means in particular that the obstacles can interpenetrate each other. This corresponds to a continuum percolation model, also called the ``Swiss cheese'' model~\cite{Hofling2006}. In fact, most of the published (simulation and theoretical) studies about continuum percolation use immobile interpenetrating obstacles~\cite{Hofling2006,Hoefling2008a,Kammerer2008,Bauer2010}.
\subsection{Immobile obstacles}
At each time step, the simulation proceeds by moving each protein independently of each other. We modeled protein diffusion as a random walk with time step $\Delta t$ and a displacement per time step $\Delta r$ that do not 
depend on the diffusion constant. Between $t$ and $t+\Delta t$, 
each protein has a probability $P_{\mathrm{move}}$ to move to a randomly-chosen 
position located at distance $\Delta r$ from its position at $t$. 
The displacement probability is given by the diffusion coefficient of the protein, $D_{\rm RW}$ : $P_{\mathrm{move}}=4D_{\rm RW}\Delta t / {\Delta r}^2$. The advantage of this algorithm is that, choosing a sufficiently small value for $\Delta r$ (namely $\Delta r < 2r_{\rm w}+2r_{\rm obs}$) ensures the excluded volume condition between proteins and obstacles for all values of $D_{\rm RW}$ and $\Delta t$. In our simulations, typical values ranged from $P_{\mathrm{move}}=0.35$ to 0.95. 
Excluded volume is then modeled by adding the restriction that
displacement attempts are rejected when the diffusing molecule, at the target 
site, overlaps with an obstacle. We have used $\Delta t=0.25 \;\mathrm{\mu s}$ and $\Delta r=1\;\mathrm{nm}$ throughout the article.

The squared displacement $R^2(t)$ of each protein was monitored 
taking into account periodic boundary conditions. Unless otherwise indicated, $R^2(t)$ was averaged over the 10 walkers across 200 initial obstacle configurations and random realizations.

\subsection{Brownian obstacles}
To model the movements of the obstacles by Brownian motion, the position of each obstacle at time $t+\Delta t$ was updated according to $\mathbf{x}_{\mathrm{obs}}(t+\Delta t)=\mathbf{x}_{\mathrm{obs}}(t)+\mathbf{N}\left(\sigma\right)$ where $\mathbf{N}\left(\sigma \right)$ is a $2$-dimensional random vector of which each component is an \textit{i.i.d.} random number with normal distribution of mean $0$ and standard deviation $\sigma$. This setting results in a diffusive movement with diffusion coefficient $D_{\mathrm{obs}}=\sigma^2/\left(2 \Delta t\right)$. When attempting to move an obstacle, if the obstacle at the chosen location is found to collide with a protein, the obstacle movement is rejected. This ensures the preservation of the excluded volume condition between obstacles and proteins. Our aim here is to specifically evaluate the hindrance caused by the obstacles on the protein movements, and not vice-versa. Therefore, we chose simulation conditions in which the hindrance caused by the proteins on the obstacle movements can be neglected. In practice, this is achieved by using only 10 diffusive proteins per simulation run. Protein motion was simulated in the same way as for immobile obstacles above.  
\subsection{OU obstacles}
The Ornstein-Uhlenbeck process can be considered a Brownian motion with additional feedback relaxation to an equilibrium position $\mu$:
\begin{equation}\label{eq:myOU}
x(t+\delta t)=x(t)+\delta t \frac{\mu-x(t)}{\tau}+\sqrt{2D_{\mathrm obs} \delta t}N
\end{equation}

where $N$ is a Gaussian random number with zero mean and unit variance, $D_{\mathrm obs}$ the diffusion constant, $\tau$ the relaxation time and $\mu$ the long-term average position (we used here $\mu = x(0)$). To simulate OU movements, we used the exact numerical simulation algorithm given in~\cite{Gillespie1996}. In our two-dimensional case it reads :
\begin{eqnarray}
\mathbf{x}_{\mathrm{obs}}(t+\Delta t)&=&\mathbf{x}_{\mathrm{obs}}(t)\exp(-\Delta t/\tau)\nonumber\\
&&+\mathbf{x}_{\mathrm{obs}}(0)\left(1-\exp(-\Delta t/\tau)\right)\nonumber\\&&+ \sqrt{D_{\mathrm obs}\tau\left(1-\exp(-2\Delta t/\tau)\right)}\mathbf{N}(1)
\end{eqnarray}
This formula is exact thence valid for all time steps $\Delta t$. Just like with Brownian obstacles, excluded volume conditions are applied between an OU obstacle and diffusing proteins, but not between two obstacles nor two proteins. Here again, protein motion was simulated in the same way as for immobile obstacles above.

\subsection{Size-distributed obstacles}
To simulate the polydispersity of the obstacles' size, we draw the radius of each obstacle as a independent Gaussian random number with mean $r_{\rm obs}$ and standard deviation $S.D.$. Negative numbers were rejected. In order keep the mean radius $r_{\rm obs}$ constant, we restricted $S.D.$ to values for which the rejections of negative-valued variates do not modify the mean radius by more than $ 0.1\%$. In practice, that means $S.D.\lesssim 1.6$ for $r_{\rm obs}=5.0$ nm.

\section{\label{sec:Results}RESULTS}
We simulated the diffusion of typical-sized proteins in two-dimensional (membrane-like) conditions, taking into account the presence of obstacles that hinder protein diffusion. To avoid numerical issues related to the size of the accessible space domain or the time-sampling, we monitored the mean-square displacement of the proteins over large time scales (at least 6 decades) with good temporal sampling ($> 700$ data points per curve) and within large-sized spatial domain ($L_{x}=L_{y}=5.0 \;\mathrm{\mu m}$). With these settings, the maximal mean  distance travelled by a protein in our simulations was 15 \% of the length of the spatial domain in 2d, thus excluding finite-size effects.

\begin{figure}[!ht]
\begin{center}
\includegraphics[scale=0.9]{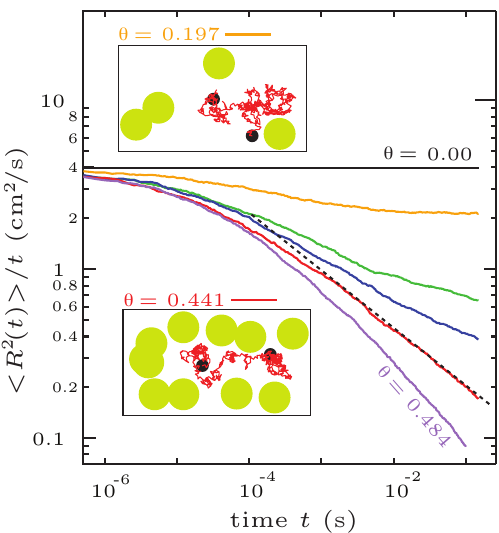}
\end{center}
\caption{(Color online) Computer simulations of subdiffusion due to immobile obstacles in two dimensions. The time-course of the ratio between the mean-squared displacement and time, \MSDt is shown for increasing obstacle densities.  Each color codes for a different obstacle density, expressed here as the excluded volume fraction $\theta$, i.e. the fraction of the accessible surface occupied by obstacles. Here, $\theta=$ 0, 0.197, 0.355, 0.395, 0.441 and 0.484 (from top to bottom). The black dashed line locates the power law $y\propto t^{-0.34}$, yielding percolation threshold $\theta_{c} = 0.441$ and anomalous diffusion exponent $\alpha=0.66$. The insets show representative trajectories (\textit{red}) of protein amidst obstacles (\textit{green} disks) at the indicated densities. The \textit{black} disks locate the initial and final position of the proteins. Parameters: protein radius $r_{\rm w}=2.0 \mathrm{\;nm}$, obstacle radius $r_{\rm obs}=5.0 \mathrm{\;nm}$, protein diffusion constant $D_{\rm RW}=1.0 \mathrm{\;\mu m^2/s}$, time step $\Delta t=0.25 \;\mathrm{\mu s}$, space step $\Delta r=1\;\mathrm{nm}$  and total domain size $L_{x}=L_{y} = 5.0 \mathrm{\;\mu m}$. All data are averages of the motion of $10$ proteins per obstacle configurations and 500 obstacle configurations.}
\label{fig:StaticObs}
\end{figure}

\subsection{Hindered diffusion by immobile obstacles}

We start with simulations of hindered diffusion with immobile obstacles. Figure~\ref{fig:StaticObs} shows the evolution with time of the rescaled mean-square displacement \MSDt 
(Fig.~\ref{fig:StaticObs}) of the diffusing proteins amidst immobile obstacles (see insets), on a log-log plot. Each curve in the figure corresponds to a different obstacle density. The top-most  curve is for unobstructed diffusion, while obstacle density 
increases from top to bottom, until the lowest curve where $48.4\%$ 
of space is occupied by obstacles (thus the excluded fraction $\theta=0.484$).

Clearly, without obstruction, \MSDt is constant which unveils a single diffusion regime, with Brownian motion ($\alpha=1$). When the excluded fraction is very high (e.g., $\theta=0.484$, well above the percolation threshold located around $\theta=0.44$, see below), the \MSDt ratio shows a supra-linear decay even at at long times, corresponding to saturation of \MSD with time. For such supra threshold obstacle densities, space is partitioned into disconnected clusters of available sites which permanently trap the proteins, thus the saturation of \MSD at long times.

For intermediate obstacle fractions ($\theta \lesssim 0.4$), three regimes can be distinguished, including two Brownian regimes: one at short times ($t\lesssim10^{-3}\mathrm{\;ms}$) that corresponds to the time for a protein to meet its first obstacle and another one at long times ($t\gtrsim0.5\mathrm{\;ms}$). Note that, within the time scale of  figure~\ref{fig:StaticObs}, the late Brownian regime is clearly reached only for the smallest obstacle densities. For larger obstacle densities, the curves display commencement of convergence to it.

According to percolation theory (that is valid for immobile obstacles) the crossover time $t_{CR}^*$ between the subdiffusive regime and the final diffusive one scales as \cite{Saxton1994}: $t_{CR}^*\propto |\theta-\theta_c |^{-z}$ where $z\approx3.8$ in two dimensions, $\theta$ is the fraction of space occupied by the obstacles and $\theta_c$ its (critical) value at the percolation threshold. This scaling, that is valid only when $\theta$ is not too far from $\theta_c$, thus predicts that $t_{CR}^*$ increases very rapidly when obstacles density approach the percolation threshold but that it is only exactly at the percolation threshold ($\theta= \theta_c$) that the system remains in the subdiffusive regime forever. This imposes a strong restriction in terms of numerical simulations: the closer to the percolation threshold, the larger the total simulation time needed in order to determine $t_CR^*$ in a precise way. In our case, we fixed the total simulation length at the largest value for which computation time remains attainable (i.e. 0.10-0.15 seconds, see Figure~\ref{fig:StaticObs}). 

The red curve, obtained with $\theta=0.441$, shows no evidence of the upward curvature typical of the crossover back to the diffusive regime (obvious with e.g. $\theta=0.395$), nor of the downward curvature typical of suprathresold obstacle densities (seen with $\theta=0.484$). We therefore consider $\theta=0.441$ as our estimate for the percolation threshold $\theta_c$. Note that albeit this is a standard way to determine the percolation threshold (see e.g.~\cite{Hofling2006}), this process can only yield an \textit{estimate} of the threshold. One generically expects that the curve for $\theta=0.441$ in Figure~\ref{fig:StaticObs} (the red curve) will eventually crossover back to the diffusive regime at times larger than the simulation length.

This estimate can be compared to theoretical predictions from continuum percolation. Expressions for the threshold in the corresponding continuum percolation problem can be found in~\cite{Bauer2010} (for $d=2$) and~\cite{Hoefling2008a} or ($d=3$) for point-like random walkers. In a first approximation, one can introduce the protein radius in these expressions by replacing the problem of a protein of size $r_{\rm w}$ within obstacles of size $r_{\rm obs}$ by that of a point-like protein diffusing amidst obstacles of size $r_{\rm obs}+r_{\rm w}$. This yields the following theoretical expressions for the critical threshold in $d=2$ dimensions:
\begin{equation}\label{eq:thetac2D}
\theta_{c}=1-\exp\left( -\pi n_{c}^{*}\left( \frac{r_{\rm obs}}{r_{\rm obs}+r_{\rm w}}\right)^{2}\right)
\end{equation}
with $n_{c}^*\approx0.359$~\cite{Bauer2010}, the critical obstacle density for point-like random walkers. For the conditions of Fig.~\ref{fig:StaticObs}, Eq.(\ref{eq:thetac2D}) yields $\theta_{c}=0.438$ in very good agreement with our estimation from the simulations (0.44). The anomalous exponent $\alpha$ can be estimated from the long-time decay of \MSDt at $\theta=\theta_{c}=0.441$ (red curve). Figure~\ref{fig:StaticObs} exhibits a clear power law decay with exponent $-0.34$, yielding the estimate $\alpha=0.66$. This value is in very good agreement with theoretical estimates from percolation  theory, $\alpha = 0.659$ in 2d~\cite{Bouchaud1990,Kammerer2008}.

\begin{figure}[!ht]
\begin{center}
\includegraphics[scale=0.9]{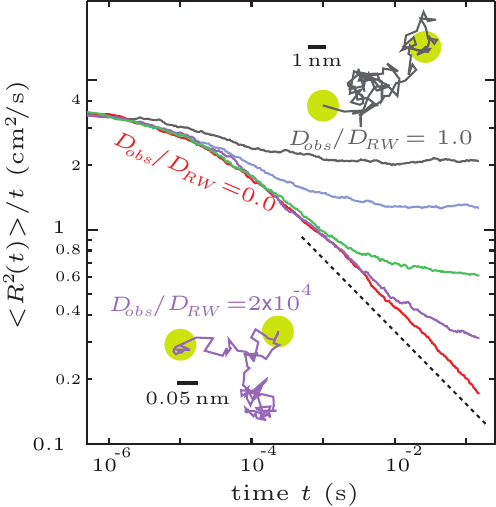}
\end{center}
\caption{(Color online) Protein diffusion in two dimensions in the presence of mobile obstacle with Brownian motion. The time-courses of  \MSDt is shown for increasing values of the diffusion constant of the obstacles $D_{\rm obs}$=0, $2\times10^{-4}$, $5\times10^{-3}$, 0.125 and 1.0 $ \mathrm{\mu m^2/s}$ (from bottom to top). The black dashed line locates the critical regime $y\propto t^{-0.34}$ for immobile obstacles. The insets show representative trajectories of the obstacles (Brownian motion) at the indicated diffusion constants. The black bars indicate the spatial scale of the trajectories and the green disks locate the starting and ending positions (the radii of the green disks are not to scale). The obstacle density was set at the percolation threshold for immobile obstacles i.e. $\theta=0.441$. Data are averaged over 10 proteins per obstacle configurations and 200 obstacle configurations. All other parameters were like in fig.~\ref{fig:StaticObs}, including the protein diffusion constant $D_{\rm RW}=1.0 \mathrm{\;\mu m^2/s}$.}
\label{fig:DiffObst}
\end{figure}

\subsection{Diffusion hindered by Brownian obstacles}
Figure~\ref{fig:DiffObst} shows simulations similar to those of Fig.~\ref{fig:StaticObs} with obstacle density at the percolation threshold ($\theta=0.414$) but where obstacles move by Brownian diffusion with diffusion constant $D_{\rm obs}$. The reference curve in this figure is the red one, that corresponds to $D_{\rm obs}=0$ i.e. immobile obstacles. The diffusion coefficient of the obstacles then progressively increases up to $D_{\rm obs}/D_{\rm RW}= 1$. The anomalous regime observed at long times with immobile obstacles (red curve) becomes first transient when obstacle start to be mobile and vanishes as soon as $D_{\rm obs}/D_{\rm RW} > 0.005$ (green curve). Hence, according to  these simulations, anomalous diffusion is not expected to persist at long times scales ( $>1 \mathrm{\;ms}$) as soon as the obstacles move with Brownian motion. Note that this point was already suggested in~\cite{Tremmel2003}, and partly in~\cite{Saxton1990}. 

However, diffusive obstacle movements are not the only possible movements for the obstacles. Because of their large size compared to that of the cell, obstacles may be restricted in their movements. Such restriction could as well be the result of stabilizing spatial interactions with each other. In other words, obstacles movement may be restricted to a confined subspace of the cell and not allowed to wander the whole cell space. This type of movement, described by an Ornstein-Uhlenbeck (OU) process, is studied in the following.

\begin{figure}[!ht]
\begin{center}
\includegraphics[scale=0.9]{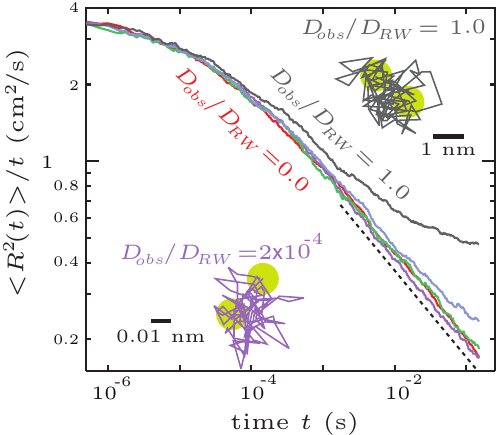}
\end{center}
\caption{(Color online) Protein diffusion in two dimensions in the presence of mobile obstacles with Ornstein-Uhlenbeck motion. The time-courses of \MSDt are shown for the same values (same color code) of the obstacle diffusion constant as in figure~\ref{fig:DiffObst}: $D_{\rm obs}$=0, $2\times10^{-4}$, $5\times10^{-3}$, 0.125 and 1.0 $ \mathrm{\mu m^2/s}$ (from bottom to top), except that the obstacle motion here is simulated by an Ornstein-Uhlenbeck process with relaxation constant $\tau=1\;\mathrm{\mu s}$. All other parameters were like in fig.~\ref{fig:DiffObst}, including the obstacle density $\theta=0.441$.}
\label{fig:OU_varD}
\end{figure}

\subsection{Diffusion hindered by Ornstein-Uhlenbeck obstacles}
Ornstein-Uhlenbeck (OU) processes are basically combination of a Brownian diffusion term with a feedback relaxation term that effectively restricts the space region explored by the obstacle (eq.~\eqref{eq:myOU}). This process has two main parameters: the diffusion constant $D_{\rm obs}$ and the time-scale $\tau$ with which the obstacle comes back close to its initial position. Figure~\ref{fig:OU_varD} shows simulations of protein diffusion amidst OU obstacles of increasing diffusion constant. The color code used for the curves is identical to that used in Fig.~\ref{fig:DiffObst} and corresponds to the same values of $D_{\rm obs}$. All other parameters are also identical to those used in Fig.~\ref{fig:DiffObst}. The only supplementary parameter, $\tau$ was set to $1\;\mathrm{\mu s}$. Comparing the representative trajectories shown in the insets of Fig.~\ref{fig:OU_varD} with those of Fig.~\ref{fig:DiffObst} (for the same obstacle diffusion constants) illustrates the fundamental difference between Brownian and OU motion. In the latter case, the obstacles are confined to a spatial region around their average position, whereas the obstacles in the Brownian case readily escape away from their initial location (this is especially true in dimension $d \geq 2$). Comparing the curves for the two types of obstacle motion reveals that  subdiffusion is much more robust with OU motion. With OU obstacles, the duration of the subdiffusive regime massively increases even for large diffusivities. For instance, when $D_{\rm obs}/D_{\rm RW}=0.125$ the anomalous regime extends over the main part of the simulation with OU motion (figure~\ref{fig:OU_varD}), whereas when the obstacles move with Brownian motion with identical value of $D_{\rm obs}/D_{\rm RW}$, the duration of the anomalous regime is several order of magnitude smaller (figure~\ref{fig:DiffObst}). 

\begin{figure}[!ht]
\begin{center}
\includegraphics[scale=0.9]{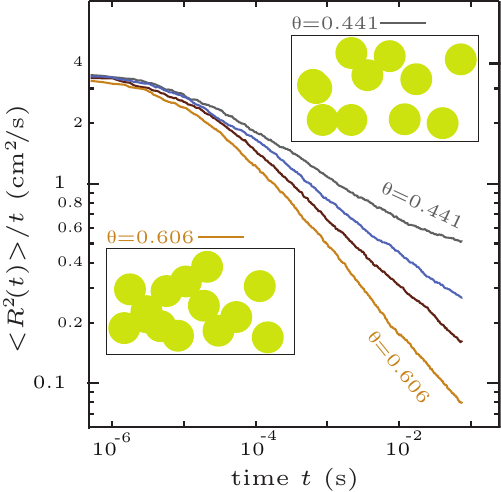}
\end{center}
\caption{(Color online) Protein diffusion amidst Ornstein-Uhlenbeck mobile obstacles at suprathreshold densities. The density of Ornstein-Uhlenbeck obstacles increases from top to bottom as $\theta=$0.441, 0.501, 0.548 and 0.606 (density is expressed as the excluded volume fraction of an equivalent number of immobile obstacles). The slopes of the \MSDt curves at long times range from 0.23 (intermediate regime for $\theta=$0.501) to 0.43 (long times for $\theta=$0.606), yielding estimates for the apparent anomalous exponents $\alpha \in [0.57-0.77]$. Insets are illustrative snapshots of obstacle locations at a time point of the simulation, intended to illustrate the difference in excluded fractions.  Parameters of the obstacle O-U motion: diffusion constant $D_{\rm obs}=1.0 \mathrm{\;\mu m^2/s}$ and relaxation constant $\tau=1\;\mathrm{\mu s}$. All other parameters were like in fig.~\ref{fig:DiffObst}, including the protein diffusion constant $D_{\rm RW}/D_{\rm obs}=1.0$.}
\label{fig:OU_varObst}
\end{figure}

\begin{figure}[!ht]	
\begin{center}
\includegraphics[scale=0.9]{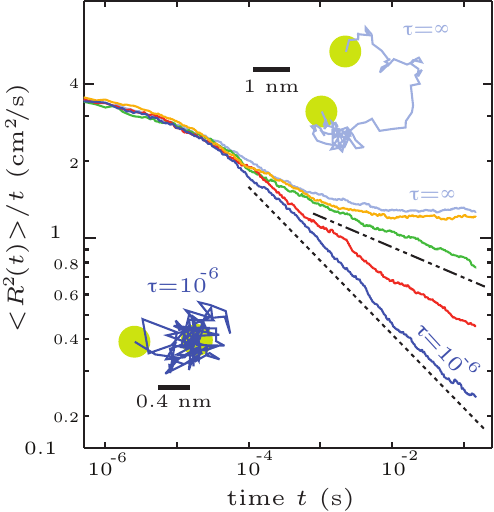}
\end{center}
\caption{(Color online) A continuum of obstacle movements from Ornstein-Uhlenbeck to Brownian motion. The time-courses \MSDt are shown for different values of the relaxation time of  OU mobile obstacles, $\tau=1,\;10,\:100$ and 1,000 $\mathrm{\mu s}$ (from bottom to top). The top-most (pale blue) curve is for Brownian motion in same conditions (formally corresponding to $\tau=\infty$). The straight lines illustrate the corresponding values of the anomalous exponent $\alpha$ when $\tau$ varies, from $\alpha=0.71$ (dashed line) to $0.88$ (dashed-dotted line). The insets show representative trajectories at the indicated values of $\tau$. The black bars indicate the spatial scale of the trajectories and the green disks locate the starting and ending positions (the radii of the green disks are not to scale). Obstacle diffusion constant $D_{\rm obs}=0.125  \mathrm{\;\mu m^2/s}$. All other parameters were like in fig.~\ref{fig:OU_varD}.}
\label{fig:OU_vartau}
\end{figure}

Another interesting property of subdiffusion in the presence of OU
obstacles is that the percolation threshold effectively disappears. In the classical case of immobile obstacles, 
the motion of proteins amidst obstacles at suprathreshold densities is limited in space since the accessible space is composed of disconnected islands of finite size. As a result, the mean-square displacement \MSD saturates at long times with supratreshold obstacle densities  (see e.g. Fig.~\ref{fig:StaticObs}, purple trace).  In two dimensions,  with immobile obstacles of radius $r_{\rm obs}=5.0  \mathrm{\;nm}$, the percolation threshold was determined above as $\theta_{c}=0.441$.  Figure~\ref{fig:OU_varObst} shows plots similar to those in 
Fig.~\ref{fig:OU_varD} except that the density of OU obstacles 
is varied above the percolation threshold of immobile obstacles $\theta > \theta_{c}$ (with $D_{\rm obs}/D_{\rm RW}=1.0$). 
The top-most curve (the gray one) corresponds to the percolation threshold. Obstacle density is then progressively increased  in the other curves of the figure, to values that would yield excluded fractions 
ranging from $\theta=0.501$ to $\theta=0.606$, if the obstacles were immobile. At first inspection, Fig.~\ref{fig:OU_varObst} shows that with OU obstacles that move as fast as the Brownian proteins, the anomalous diffusion regime is preserved for all curves, i.e. even when the obstacle density is much larger than the percolation threshold. Moreover, the \MSDt plots for suprathreshold conditions do not display the supralinear decay typical of protein diffusion amidst suprathreshold immobile obstacles observed Fig.~\ref{fig:StaticObs}. 

Above the threshold, examination of the plots indicates convincing power law decays even at long time scales, with exponents that vary from 0.57 to 0.77. Therefore, for large mobile obstacle densities, our simulations suggest the existence of power-law regimes with exponents that are significantly smaller than the universal value of percolation theory (0.659). This range is in fact compatible with most experimentally determined values of $\alpha$ \textit{in vivo} \cite{Hoefling2013}. Therefore, protein diffusion hindered by OU obstacles above the threshold could be one explanation for the reported variations in the values of $\alpha$.

Whether or not the power law regimes observed for suprathreshold obstacle densities are permanent or would exhibit a crossover back to the diffusive regime at time larger than the total simulation length in the figure, cannot be decided on the basis of our results. Indeed, with a power-law function of time, improvement in precision is expected only for simulation lengths that would be at least ten-fold larger than on figure~\ref{fig:OU_varObst}. The computation cost of such simulations are however prohibitive (especially at such large density of mobile obstacles). However, we do not think that our results can be explained by a potential crossover back to diffusion. The crossover back to the diffusive regime can lead to an effective (or apparent) power-law, with an exponent that would be between the universal value of the anomalous exponent (0.659) and 1. In the cases of figure~\ref{fig:OU_varObst} however, the exponents obtained for the largest obstacle densities are significantly smaller than 0.659 (down to 0.57), which cannot be accounted for by the potential transient nature of the phenomenon. We therefore suspect they reflect a more fundamental property of obstructed random walks with mobile obstacles moving according to an O-U process.

\begin{figure}[!t]
\begin{center}
\includegraphics[scale=0.9]{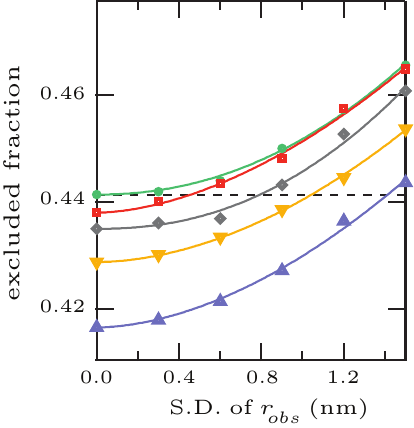}
\end{center}
\caption{(Color online) The excluded fraction depends on the variance of the obstacle size. Each curve shows the evolution of the excluded fraction (for immobile obstacles) with the standard deviation S.D. of the obstacle radius. Obstacle size was drawn from a Normal distribution with mean 5.0 nm and variance $\left(S.D.\right)^{2}$. The number of randomly located obstacles was (from bottom to top) $172.10^{3},\;177.10^{3},\;182.10^{3},\;183.5.10^{3}\;\mathrm{and}\;185.10^{3}$. The dashed line locates the excluded fraction at the percolation threshold (0.441).}
\label{fig:OU_varSD_Frac}
\end{figure}

\begin{figure}[!t]
\begin{center}
\includegraphics[scale=0.9]{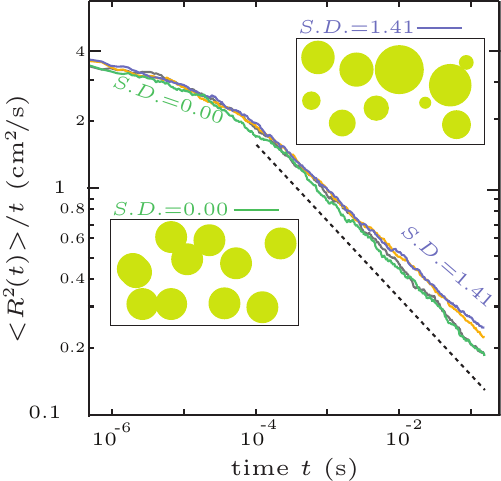}
\end{center}
\caption{(Color online) Protein diffusion amidst polydisperse Ornstein-Uhlenbeck mobile obstacles. The size of the mobile obstacles was a random variable drawn according to Normal distribution with mean $5.0 \mathrm{\;nm}$ and standard deviation (S.D.) = 0, 0.45, 0.80, 1.04 and 1.41 nm (from bottom to top). Parameters of the OU motion for the obstacles $D_{\rm obs}=5\times 10^{-3} \mathrm{\;\mu m^2/s}$, $\tau=1\;\mathrm{\mu s}$.
For each value of the standard deviation, the total number of obstacles was adjusted so that the excluded fraction was kept to the critical threshold of immobile obstacles $\theta=0.441$ (see text and fig.~\ref{fig:OU_varSD_Frac}). All other parameters were like in fig.~\ref{fig:OU_varD}.}
\label{fig:OU_varSD}
\end{figure}

Investigating further the effects of OU obstacles, we uncovered another possible explanation for the experimental reports of variable $\alpha$ values. Formally, the relaxation time scale $\tau$ allows to go continuously from a OU motion ($\tau\to 0$) to a Brownian motion ($\tau \to \infty$). In Figure~\ref{fig:OU_vartau}, the relaxation time scale $\tau$ was varied from $1 \;\mathrm{\mu s}$ to $1 \;\mathrm{ms}$. For short relaxations, the results of the previous section are regained, showing an subdiffusion regime with $\alpha=0.71$. Increasing $\tau$ above $1.0 \;\mathrm{ms}$ ultimately produces the same results as protein diffusion in Brownian diffusion, i.e. an almost complete disappearance of the anomalous region. Nevertheless, between those two extremes, protein diffusion appears to preserve the anomalous regime even for long times, but with exponent $\alpha$ that varies with $\tau$ (ranging from 0.71 to 0.88 in the figure). Albeit these power-law regimes may  not be genuine power laws but reflect the crossover regime back to diffusive motion, they could account for the variations of $\alpha$ measured \textit{in vivo}.

\subsection{Effects of obstacle polydispersity}
Another frequent simplification made in computer simulations of diffusion hindered by obstacles resides in the variability of the obstacle size. In most studies, the obstacle size is monodisperse, i.e. all obstacles in a given simulation condition have the same radius. We next investigated whether random obstacle sizes could have an effect on the diffusion of the proteins. To this end, we ran the same simulations as in Figure~\ref{fig:OU_varD}., for instance, except that the radius of each obstacle is no more set to a constant value $r_{\rm obs}$, but is drawn from a Gaussian distribution with mean $r_{\rm obs}$ and standard deviation $S.D.$. Note that when the variance of the obstacle radius increases, the excluded fraction for identical numbers of immobile obstacles increases too (Figure~\ref{fig:OU_varSD_Frac}). Therefore, we had to adjust the number of obstacles to keep the excluded fraction constant with increasing size variance. Figure~\ref{fig:OU_varSD} shows the results obtained when the variance of the obstacle radius increases for OU obstacles. It seems clear from these simulations that the polydispersity of the obstacle sizes does \textit{not} have a strong influence on the diffusion regime of the proteins, except for the largest variances tested where the anomalous regime tends to deviate a bit from a clear power law (at long time scales). We thus conclude that the broadness of the obstacle size distribution is not likely to have a strong influence of the value on the anomalous exponent $\alpha$ in cells.

\section{\label{sec:Discussion}DISCUSSION}

%The simulations reported here first of all confirm that 
%hindered diffusion due to randomly-located immobile obstacles does not feature 
%asymptotic anomalous scaling regimes where the exponent value would decay with obstacle density. 
%The anomalous exponent is unique, only observable very near the percolation 
%threshold, and obstacle density only modifies the duration of the subdiffusion regime. This is in strong contrast to numerous experimental studies, 
%in vitro~\cite{Weiss2004,Banks2005,Szymanski2009, Horton2010} as well as 
%in vivo~\cite{Schwille1999,Smith1999,Wachsmuth2000,Platani2002,Murase2004,Weiss2004,Golding2006,Guigas2007}, 
%that have consistently reported anomalous diffusion regimes with variable values of the 
%exponent $\alpha$.  In the present article, we have investigated wether this discrepancy could be resolved by taking into account the motion of the obstacles. 

Our simulations confirm the conclusion of previous studies that the 
long-time anomalous regime typical of immobile obstacles disappears very rapidly 
as soon as the obstacles are mobile. However, our finding that diffusion amidst Ornstein-Uhlenbeck (OU) mobile obstacles gives rise to extended anomalous regimes has interesting implications. For instance, light-harvesting complexes of photosynthetic membranes are 
large-size obstacles that occupy between 70 and 90\% of the membrane 
area~\cite{Tremmel2003}.  Recent experimental observations revealed that their 
mobility in the membrane indeed consists of fluctuations around an equilibrium 
position, with fluctuation amplitudes that depend on the considered region in 
the membrane~\cite{Scheuring2006}. This seems a good example of an OU movement. 

Our results are also relevant to diffusion in lipid membranes and in particular in binary lipid membranes close to the fluid/gel critical point~\cite{Ehrig2011}. In these systems, the gel phase is made of dynamically-rearranging, fluctuating and interpenetrating domains dispersed in the fluid phase. Ehrig {\it et al.} (2011) \cite{Ehrig2011} showed that the diffusion of marker lipids that are restricted to the fluid phase of these systems is transiently anomalous with anomalous diffusion profiles that are qualitatively similar to ours (compare eg their Figures 4a,b with our Fig.~\ref{fig:StaticObs}). Precise quantitative comparisons are impossible because our simulations concern protein diffusion while~\cite{Ehrig2011} considered the diffusion of lipids, that is several order of magnitude faster. Moreover, in the simulations shown in~\cite{Ehrig2011} (with no additional elements such as interactions with the cytoskeleton), the amplitude of the anomalous regime is limited. We thus lack the necessary ingredients to judge of the agreement with our paper in strongly anomalous cases. Nevertheless, qualitatively, the agreement between the behaviors we obtained using OU obstacles with large mobility or small density and their results is good. Hence, our model can be considered a coarse-grained approach where the gel domains are directly modeled as disk-shaped obstacles and the lipid molecules of the membrane are not explicitly represented. This coarse-grained model greatly facilitates exploration of the parameter space in particular concerning the size of the gel domains/obstacles and/or their mobility.

%Our results with OU obstacles could also explain 
%some of the experimental reports obtained in vitro for tracers diffusing in 
%concentrated mobile crowders. Horton {\it et al.} (2010) \cite{Horton2010} for instance, studied 2d 
%diffusion on artificial lipid bilayer membranes with avidin as both the crowder 
%and the tracer. This study showed that even at small densities, avidin self-organizes 
%into a spatially heterogeneous (possibly gel-like) structure, that persists 
%(i.e. does not mix) for several tens of minutes. This type of spatial heterogeneity 
%is typically expected to yield OU obstacles. In agreement with our 
%simulation results for OU obstacles (Fig.~\ref{fig:OU_varD}), the authors 
%find clear evidence for transient anomalous diffusion. 

One major conclusion from our work is that proteins diffusing amidst hindering obstacles may undergo subdiffusion with properties that are very similar to those measured \textit{in vivo}, whenever the mobile obstacle motion is an OU process. Of particular importance, we observed that the subdiffusive motion of the tracked protein displays convincing power-laws with anomalous exponent $\alpha$ that varies with the density of OU obstacles (above the percolation threshold of immobile obstacles) or the relaxation time-scale of the OU process. In particular, we observed values (e.g. $\alpha=0.57$, Fig.~\ref{fig:OU_varObst}) that are significantly below the universal value of $\alpha=0.659$ predicted in 2d in the case of immobile obstacles. Therefore, subdiffusion due to mobile obstacles with OU-type of motion may account for the large variation range exhibited by experimental measurements of $\alpha$ in living cells (see Section \ref{sec:intro}) and the fact that some of these experimental estimates in 2d \cite{Smith1999,Murase2004} are below the universal value predicted for immobile obstacles in 2d by percolation theory.

In our simulations, the time of crossover from the anomalous regime back to the diffusive one increases rapidly when obstacle concentration increases. Of course, crossover times larger than the total simulation length, that is of the order of 0.1 second, could not be observed. Globally, the first crossover from the initial Brownian regime to the subdiffusive one in our work takes place at around 0.01 to 0.10 milliseconds and the subdiffusive regime has a duration that varies from some milliseconds to more than 100 milliseconds. In experimental reports, the time scales of the observed anomalous regimes are spread over several orders of magnitude. In bacteria, for instance, the diffusion of large objects (ribosomes or entire chromosome loci) displays anomalous diffusion regimes that usually last for very long time scales, up to 10 or even 100 second \cite{Golding2006,Weber2010}, although hindered diffusion due to obstacles might not be the cause of anomalous subdiffusion in these cases (see e.g.~\cite{Weber2010}). On the other hand, in eukaryotic cells, the anomalous regime is often observed with time scales of 0.1 to 100 milliseconds\cite{Schwille1999,Wachsmuth2000, Jeon2011, Weiss2004, Guigas2007, Guigas2007a, Murase2004}, which is the same time scale as in our simulations. Therefore, we think our work may constitute a potential explanation, at least for these experimental situations. In many cases though, the time scale of the measured anomalous regime is much larger, e.g. from 1 to 100 seconds \cite{Smith1999, Bronstein2009, Tabei2013, Feder1996, Vrljic2002} even up to several hundred minutes~\cite{Platani2002}. However the microscopic origin of these very long-time scale anomalous regimes is often complex, possibly combining different sources of anomalous transport (see e.g.~\cite{Weigel2011,Tabei2013}). These situations cannot be accounted for with our simulations that only account for anomalous diffusion due to hindering by obstacles.  

In the experimental recordings where the transition from the subdiffusive regime back to the diffusive one was observed, the corresponding crossover times ranged between 0.1 seconds~\cite{Murase2004} and 100 seconds~\cite{Bronstein2009} or even 100 minutes~\cite{Platani2002}. Even after factoring out the fact that distinct cell types or intracellular environments can experience distinct obstacle densities~\cite{Kuhn2011}, our work can hardly account for the totality of this substantial interval of time scales (more than 4 orders of magnitude). However, our results are clearly compatible with the shorter time scales reported.

The OU process used here is mostly used to preclude the escape of the mobile obstacles too far from their initial (equilibrium) positions. Other processes with this property could be considered. In space dimension $d \geq 2$, the Brownian motion is a non-compact exploration process, i.e. the random walker visits only a small part of the available space (or volume). As a result the probability that a random walker escapes its initial position (never coming back to its initial location, at any time) is finite but non zero~\cite{Gennes1983}. By contrast, in $d < 2$ (e.g. in one space dimension), a random walker experiencing Brownian motion comes back to its initial position almost surely, i.e. its escape probability vanishes. One-dimensional Brownian motion is a compact exploration process. Brownian obstacles in our two-dimensional simulations thus tend to wander away from their initial positions. This is clearly associated with a rapid disappearance of the anomalous diffusion regime of the proteins that move amidst them. To change the tendencies of the Brownian obstacles to escape their position, we have introduced here obstacles that move via an Ornstein-Uhlenbeck process. The OU process is however not the only way by which one may reduce the probability that moving obstacles escape their initial position. Another possibility could be that the obstacles themselves undergo subdiffusion. Indeed, subdiffusion is a compact exploration process of the space around the walker~\cite{Condamin2007}. Therefore, obstacles moving by an subdiffusion process may hinder the diffusion of smaller proteins in a similar way as observed here for OU obstacles. This enticing possibility would correspond to simulating the subdiffusion of proteins due to hindering by obstacles themselves undergoing subdiffusion on another time scale. Albeit challenging in methodological terms, this approach may allow to decipher several of the remaining issues uncovered by the experimental measurements of biomolecule diffusion \textit{in vivo}.

\section{Acknowledgments}
The authors acknowledge the support of the computing centre of CNRS IN2P3 in Lyon (http://cc.in2p3.fr), where the simulations were performed. This research was funded by the French National Institute for Research in Computer Science and Control (INRIA, grant ``Action d'Envergure ColAge'') and the French National Agency for Research (ANR, grant ``PAGDEG ANR-09-PIRI-0030'').

\bibliography{BerryChate_Revised}

%merlin.mbs apsrev4-1.bst 2010-07-25 4.21a (PWD, AO, DPC) hacked
%Control: key (0)
%Control: author (8) initials jnrlst
%Control: editor formatted (1) identically to author
%Control: production of article title (-1) disabled
%Control: page (0) single
%Control: year (1) truncated
%Control: production of eprint (0) enabled
\begin{thebibliography}{50}%
\makeatletter
\providecommand \@ifxundefined [1]{%
 \@ifx{#1\undefined}
}%
\providecommand \@ifnum [1]{%
 \ifnum #1\expandafter \@firstoftwo
 \else \expandafter \@secondoftwo
 \fi
}%
\providecommand \@ifx [1]{%
 \ifx #1\expandafter \@firstoftwo
 \else \expandafter \@secondoftwo
 \fi
}%
\providecommand \natexlab [1]{#1}%
\providecommand \enquote  [1]{``#1''}%
\providecommand \bibnamefont  [1]{#1}%
\providecommand \bibfnamefont [1]{#1}%
\providecommand \citenamefont [1]{#1}%
\providecommand \href@noop [0]{\@secondoftwo}%
\providecommand \href [0]{\begingroup \@sanitize@url \@href}%
\providecommand \@href[1]{\@@startlink{#1}\@@href}%
\providecommand \@@href[1]{\endgroup#1\@@endlink}%
\providecommand \@sanitize@url [0]{\catcode `\\12\catcode `\$12\catcode
  `\&12\catcode `\#12\catcode `\^12\catcode `\_12\catcode `\%12\relax}%
\providecommand \@@startlink[1]{}%
\providecommand \@@endlink[0]{}%
\providecommand \url  [0]{\begingroup\@sanitize@url \@url }%
\providecommand \@url [1]{\endgroup\@href {#1}{\urlprefix }}%
\providecommand \urlprefix  [0]{URL }%
\providecommand \Eprint [0]{\href }%
\providecommand \doibase [0]{http://dx.doi.org/}%
\providecommand \selectlanguage [0]{\@gobble}%
\providecommand \bibinfo  [0]{\@secondoftwo}%
\providecommand \bibfield  [0]{\@secondoftwo}%
\providecommand \translation [1]{[#1]}%
\providecommand \BibitemOpen [0]{}%
\providecommand \bibitemStop [0]{}%
\providecommand \bibitemNoStop [0]{.\EOS\space}%
\providecommand \EOS [0]{\spacefactor3000\relax}%
\providecommand \BibitemShut  [1]{\csname bibitem#1\endcsname}%
\let\auto@bib@innerbib\@empty
%</preamble>
\bibitem [{\citenamefont {Schwille}\ \emph {et~al.}(1999)\citenamefont
  {Schwille}, \citenamefont {Korlach},\ and\ \citenamefont
  {Webb}}]{Schwille1999}%
  \BibitemOpen
  \bibfield  {author} {\bibinfo {author} {\bibfnamefont {P.}~\bibnamefont
  {Schwille}}, \bibinfo {author} {\bibfnamefont {J.}~\bibnamefont {Korlach}}, \
  and\ \bibinfo {author} {\bibfnamefont {W.}~\bibnamefont {Webb}},\ }\href@noop
  {} {\bibfield  {journal} {\bibinfo  {journal} {Cytometry}\ }\textbf {\bibinfo
  {volume} {36, 176-182}} (\bibinfo {year} {1999})}\BibitemShut {NoStop}%
\bibitem [{\citenamefont {Smith}\ \emph {et~al.}(1999)\citenamefont {Smith},
  \citenamefont {Morrison}, \citenamefont {Wilson}, \citenamefont
  {Fernández},\ and\ \citenamefont {Cherry}}]{Smith1999}%
  \BibitemOpen
  \bibfield  {author} {\bibinfo {author} {\bibfnamefont {P.~R.}\ \bibnamefont
  {Smith}}, \bibinfo {author} {\bibfnamefont {I.~E.}\ \bibnamefont {Morrison}},
  \bibinfo {author} {\bibfnamefont {K.~M.}\ \bibnamefont {Wilson}}, \bibinfo
  {author} {\bibfnamefont {N.}~\bibnamefont {Fernández}}, \ and\ \bibinfo
  {author} {\bibfnamefont {R.~J.}\ \bibnamefont {Cherry}},\ }\href {\doibase
  10.1016/S0006-3495(99)77486-2} {\bibfield  {journal} {\bibinfo  {journal}
  {Biophys J}\ }\textbf {\bibinfo {volume} {76}},\ \bibinfo {pages} {3331}
  (\bibinfo {year} {1999})}\BibitemShut {NoStop}%
\bibitem [{\citenamefont {Wachsmuth}\ \emph {et~al.}(2000)\citenamefont
  {Wachsmuth}, \citenamefont {Waldeck},\ and\ \citenamefont
  {Langowski}}]{Wachsmuth2000}%
  \BibitemOpen
  \bibfield  {author} {\bibinfo {author} {\bibfnamefont {M.}~\bibnamefont
  {Wachsmuth}}, \bibinfo {author} {\bibfnamefont {W.}~\bibnamefont {Waldeck}},
  \ and\ \bibinfo {author} {\bibfnamefont {J.}~\bibnamefont {Langowski}},\
  }\href@noop {} {\bibfield  {journal} {\bibinfo  {journal} {J. Mol. Biol.}\
  }\textbf {\bibinfo {volume} {298, 677-689}} (\bibinfo {year}
  {2000})}\BibitemShut {NoStop}%
\bibitem [{\citenamefont {Seisenberger}\ \emph {et~al.}(2001)\citenamefont
  {Seisenberger}, \citenamefont {Ried}, \citenamefont {Endress}, \citenamefont
  {Büning}, \citenamefont {Hallek},\ and\ \citenamefont
  {Bräuchle}}]{Seisenberger2001}%
  \BibitemOpen
  \bibfield  {author} {\bibinfo {author} {\bibfnamefont {G.}~\bibnamefont
  {Seisenberger}}, \bibinfo {author} {\bibfnamefont {M.~U.}\ \bibnamefont
  {Ried}}, \bibinfo {author} {\bibfnamefont {T.}~\bibnamefont {Endress}},
  \bibinfo {author} {\bibfnamefont {H.}~\bibnamefont {Büning}}, \bibinfo
  {author} {\bibfnamefont {M.}~\bibnamefont {Hallek}}, \ and\ \bibinfo {author}
  {\bibfnamefont {C.}~\bibnamefont {Bräuchle}},\ }\href {\doibase
  10.1126/science.1064103} {\bibfield  {journal} {\bibinfo  {journal}
  {Science}\ }\textbf {\bibinfo {volume} {294}},\ \bibinfo {pages} {1929}
  (\bibinfo {year} {2001})}\BibitemShut {NoStop}%
\bibitem [{\citenamefont {Fujiwara}\ \emph {et~al.}(2002)\citenamefont
  {Fujiwara}, \citenamefont {Ritchie}, \citenamefont {Murakoshi}, \citenamefont
  {Jacobson},\ and\ \citenamefont {Kusumi}}]{Fujiwara2002}%
  \BibitemOpen
  \bibfield  {author} {\bibinfo {author} {\bibfnamefont {T.}~\bibnamefont
  {Fujiwara}}, \bibinfo {author} {\bibfnamefont {K.}~\bibnamefont {Ritchie}},
  \bibinfo {author} {\bibfnamefont {H.}~\bibnamefont {Murakoshi}}, \bibinfo
  {author} {\bibfnamefont {K.}~\bibnamefont {Jacobson}}, \ and\ \bibinfo
  {author} {\bibfnamefont {A.}~\bibnamefont {Kusumi}},\ }\href {\doibase
  10.1083/jcb.200202050} {\bibfield  {journal} {\bibinfo  {journal} {J Cell
  Biol}\ }\textbf {\bibinfo {volume} {157}},\ \bibinfo {pages} {1071} (\bibinfo
  {year} {2002})}\BibitemShut {NoStop}%
\bibitem [{\citenamefont {Caspi}\ \emph {et~al.}(2002)\citenamefont {Caspi},
  \citenamefont {Granek},\ and\ \citenamefont {Elbaum}}]{Caspi2002}%
  \BibitemOpen
  \bibfield  {author} {\bibinfo {author} {\bibfnamefont {A.}~\bibnamefont
  {Caspi}}, \bibinfo {author} {\bibfnamefont {R.}~\bibnamefont {Granek}}, \
  and\ \bibinfo {author} {\bibfnamefont {M.}~\bibnamefont {Elbaum}},\
  }\href@noop {} {\bibfield  {journal} {\bibinfo  {journal} {Phys. Rev. E}\
  }\textbf {\bibinfo {volume} {66, 011916}} (\bibinfo {year}
  {2002})}\BibitemShut {NoStop}%
\bibitem [{\citenamefont {Platani}\ \emph {et~al.}(2002)\citenamefont
  {Platani}, \citenamefont {Goldberg}, \citenamefont {Lamond},\ and\
  \citenamefont {Swedlow}}]{Platani2002}%
  \BibitemOpen
  \bibfield  {author} {\bibinfo {author} {\bibfnamefont {M.}~\bibnamefont
  {Platani}}, \bibinfo {author} {\bibfnamefont {I.}~\bibnamefont {Goldberg}},
  \bibinfo {author} {\bibfnamefont {A.~I.}\ \bibnamefont {Lamond}}, \ and\
  \bibinfo {author} {\bibfnamefont {J.~R.}\ \bibnamefont {Swedlow}},\ }\href
  {\doibase 10.1038/ncb809} {\bibfield  {journal} {\bibinfo  {journal} {Nat
  Cell Biol}\ }\textbf {\bibinfo {volume} {4}},\ \bibinfo {pages} {502}
  (\bibinfo {year} {2002})}\BibitemShut {NoStop}%
\bibitem [{\citenamefont {Tolic-Norrelykke}\ \emph {et~al.}(2004)\citenamefont
  {Tolic-Norrelykke}, \citenamefont {Munteanu}, \citenamefont {Thon},
  \citenamefont {Oddershede},\ and\ \citenamefont
  {Berg-Sorensen}}]{Tolic-Noerrelykke2004}%
  \BibitemOpen
  \bibfield  {author} {\bibinfo {author} {\bibfnamefont {I.~M.}\ \bibnamefont
  {Tolic-Norrelykke}}, \bibinfo {author} {\bibfnamefont {E.-L.}\ \bibnamefont
  {Munteanu}}, \bibinfo {author} {\bibfnamefont {G.}~\bibnamefont {Thon}},
  \bibinfo {author} {\bibfnamefont {L.}~\bibnamefont {Oddershede}}, \ and\
  \bibinfo {author} {\bibfnamefont {K.}~\bibnamefont {Berg-Sorensen}},\
  }\href@noop {} {\bibfield  {journal} {\bibinfo  {journal} {Phys. Rev. Lett.}\
  }\textbf {\bibinfo {volume} {93}},\ \bibinfo {pages} {078102} (\bibinfo
  {year} {2004})}\BibitemShut {NoStop}%
\bibitem [{\citenamefont {Golding}\ and\ \citenamefont
  {Cox}(2006)}]{Golding2006}%
  \BibitemOpen
  \bibfield  {author} {\bibinfo {author} {\bibfnamefont {I.}~\bibnamefont
  {Golding}}\ and\ \bibinfo {author} {\bibfnamefont {E.~C.}\ \bibnamefont
  {Cox}},\ }\href@noop {} {\bibfield  {journal} {\bibinfo  {journal} {Phys.
  Rev. Lett.}\ }\textbf {\bibinfo {volume} {96}},\ \bibinfo {pages} {098102}
  (\bibinfo {year} {2006})}\BibitemShut {NoStop}%
\bibitem [{\citenamefont {Bronstein}\ \emph {et~al.}(2009)\citenamefont
  {Bronstein}, \citenamefont {Israel}, \citenamefont {Kepten}, \citenamefont
  {Mai}, \citenamefont {Shav-Tal}, \citenamefont {Barkai},\ and\ \citenamefont
  {Garini}}]{Bronstein2009}%
  \BibitemOpen
  \bibfield  {author} {\bibinfo {author} {\bibfnamefont {I.}~\bibnamefont
  {Bronstein}}, \bibinfo {author} {\bibfnamefont {Y.}~\bibnamefont {Israel}},
  \bibinfo {author} {\bibfnamefont {E.}~\bibnamefont {Kepten}}, \bibinfo
  {author} {\bibfnamefont {S.}~\bibnamefont {Mai}}, \bibinfo {author}
  {\bibfnamefont {Y.}~\bibnamefont {Shav-Tal}}, \bibinfo {author}
  {\bibfnamefont {E.}~\bibnamefont {Barkai}}, \ and\ \bibinfo {author}
  {\bibfnamefont {Y.}~\bibnamefont {Garini}},\ }\href {\doibase
  10.1103/PhysRevLett.103.018102} {\bibfield  {journal} {\bibinfo  {journal}
  {Phys. Rev. Lett.}\ }\textbf {\bibinfo {volume} {103}},\ \bibinfo {pages}
  {018102} (\bibinfo {year} {2009})}\BibitemShut {NoStop}%
\bibitem [{\citenamefont {Weber}\ \emph {et~al.}(2010)\citenamefont {Weber},
  \citenamefont {Spakowitz},\ and\ \citenamefont {Theriot}}]{Weber2010}%
  \BibitemOpen
  \bibfield  {author} {\bibinfo {author} {\bibfnamefont {S.~C.}\ \bibnamefont
  {Weber}}, \bibinfo {author} {\bibfnamefont {A.~J.}\ \bibnamefont
  {Spakowitz}}, \ and\ \bibinfo {author} {\bibfnamefont {J.~A.}\ \bibnamefont
  {Theriot}},\ }\href@noop {} {\bibfield  {journal} {\bibinfo  {journal} {Phys
  Rev Lett}\ }\textbf {\bibinfo {volume} {104}},\ \bibinfo {pages} {238102}
  (\bibinfo {year} {2010})}\BibitemShut {NoStop}%
\bibitem [{\citenamefont {Jeon}\ \emph {et~al.}(2011)\citenamefont {Jeon},
  \citenamefont {Tejedor}, \citenamefont {Burov}, \citenamefont {Barkai},
  \citenamefont {Selhuber-Unkel}, \citenamefont {Berg-S\o{}rensen},
  \citenamefont {Oddershede},\ and\ \citenamefont {Metzler}}]{Jeon2011}%
  \BibitemOpen
  \bibfield  {author} {\bibinfo {author} {\bibfnamefont {J.-H.}\ \bibnamefont
  {Jeon}}, \bibinfo {author} {\bibfnamefont {V.}~\bibnamefont {Tejedor}},
  \bibinfo {author} {\bibfnamefont {S.}~\bibnamefont {Burov}}, \bibinfo
  {author} {\bibfnamefont {E.}~\bibnamefont {Barkai}}, \bibinfo {author}
  {\bibfnamefont {C.}~\bibnamefont {Selhuber-Unkel}}, \bibinfo {author}
  {\bibfnamefont {K.}~\bibnamefont {Berg-S\o{}rensen}}, \bibinfo {author}
  {\bibfnamefont {L.}~\bibnamefont {Oddershede}}, \ and\ \bibinfo {author}
  {\bibfnamefont {R.}~\bibnamefont {Metzler}},\ }\href {\doibase
  10.1103/PhysRevLett.106.048103} {\bibfield  {journal} {\bibinfo  {journal}
  {Phys. Rev. Lett.}\ }\textbf {\bibinfo {volume} {106}},\ \bibinfo {pages}
  {048103} (\bibinfo {year} {2011})}\BibitemShut {NoStop}%
\bibitem [{\citenamefont {Weigel}\ \emph {et~al.}(2011)\citenamefont {Weigel},
  \citenamefont {Simon}, \citenamefont {Tamkun},\ and\ \citenamefont
  {Krapf}}]{Weigel2011}%
  \BibitemOpen
  \bibfield  {author} {\bibinfo {author} {\bibfnamefont {A.~V.}\ \bibnamefont
  {Weigel}}, \bibinfo {author} {\bibfnamefont {B.}~\bibnamefont {Simon}},
  \bibinfo {author} {\bibfnamefont {M.~M.}\ \bibnamefont {Tamkun}}, \ and\
  \bibinfo {author} {\bibfnamefont {D.}~\bibnamefont {Krapf}},\ }\href
  {\doibase 10.1073/pnas.1016325108} {\bibfield  {journal} {\bibinfo  {journal}
  {Proc Natl Acad Sci U S A}\ }\textbf {\bibinfo {volume} {108}},\ \bibinfo
  {pages} {6438} (\bibinfo {year} {2011})}\BibitemShut {NoStop}%
\bibitem [{\citenamefont {Tabei}\ \emph {et~al.}(2013)\citenamefont {Tabei},
  \citenamefont {Burov}, \citenamefont {Kim}, \citenamefont {Kuznetsov},
  \citenamefont {Huynh}, \citenamefont {Jureller}, \citenamefont {Philipson},
  \citenamefont {Dinner},\ and\ \citenamefont {Scherer}}]{Tabei2013}%
  \BibitemOpen
  \bibfield  {author} {\bibinfo {author} {\bibfnamefont {S.~M.~A.}\
  \bibnamefont {Tabei}}, \bibinfo {author} {\bibfnamefont {S.}~\bibnamefont
  {Burov}}, \bibinfo {author} {\bibfnamefont {H.~Y.}\ \bibnamefont {Kim}},
  \bibinfo {author} {\bibfnamefont {A.}~\bibnamefont {Kuznetsov}}, \bibinfo
  {author} {\bibfnamefont {T.}~\bibnamefont {Huynh}}, \bibinfo {author}
  {\bibfnamefont {J.}~\bibnamefont {Jureller}}, \bibinfo {author}
  {\bibfnamefont {L.~H.}\ \bibnamefont {Philipson}}, \bibinfo {author}
  {\bibfnamefont {A.~R.}\ \bibnamefont {Dinner}}, \ and\ \bibinfo {author}
  {\bibfnamefont {N.~F.}\ \bibnamefont {Scherer}},\ }\href {\doibase
  10.1073/pnas.1221962110} {\bibfield  {journal} {\bibinfo  {journal} {Proc
  Natl Acad Sci U S A}\ }\textbf {\bibinfo {volume} {110}},\ \bibinfo {pages}
  {4911} (\bibinfo {year} {2013})}\BibitemShut {NoStop}%
\bibitem [{\citenamefont {Hoefling}\ and\ \citenamefont
  {Franosch}(2013)}]{Hoefling2013}%
  \BibitemOpen
  \bibfield  {author} {\bibinfo {author} {\bibfnamefont {F.}~\bibnamefont
  {Hoefling}}\ and\ \bibinfo {author} {\bibfnamefont {T.}~\bibnamefont
  {Franosch}},\ }\href {\doibase 10.1088/0034-4885/76/4/046602} {\bibfield
  {journal} {\bibinfo  {journal} {Rep Prog Phys}\ }\textbf {\bibinfo {volume}
  {76}},\ \bibinfo {pages} {046602} (\bibinfo {year} {2013})}\BibitemShut
  {NoStop}%
\bibitem [{\citenamefont {Kulic}\ \emph {et~al.}(2008)\citenamefont {Kulic},
  \citenamefont {Brown}, \citenamefont {Kim}, \citenamefont {Kural},
  \citenamefont {Blehm}, \citenamefont {Selvin}, \citenamefont {Nelson},\ and\
  \citenamefont {Gelfand}}]{Kulic2008}%
  \BibitemOpen
  \bibfield  {author} {\bibinfo {author} {\bibfnamefont {I.~M.}\ \bibnamefont
  {Kulic}}, \bibinfo {author} {\bibfnamefont {A.~E.~X.}\ \bibnamefont {Brown}},
  \bibinfo {author} {\bibfnamefont {H.}~\bibnamefont {Kim}}, \bibinfo {author}
  {\bibfnamefont {C.}~\bibnamefont {Kural}}, \bibinfo {author} {\bibfnamefont
  {B.}~\bibnamefont {Blehm}}, \bibinfo {author} {\bibfnamefont {P.~R.}\
  \bibnamefont {Selvin}}, \bibinfo {author} {\bibfnamefont {P.~C.}\
  \bibnamefont {Nelson}}, \ and\ \bibinfo {author} {\bibfnamefont {V.~I.}\
  \bibnamefont {Gelfand}},\ }\href@noop {} {\bibfield  {journal} {\bibinfo
  {journal} {Proc. Natl. Acad. Sci. USA}\ }\textbf {\bibinfo {volume} {105}},\
  \bibinfo {pages} {10011} (\bibinfo {year} {2008})}\BibitemShut {NoStop}%
\bibitem [{\citenamefont {Elowitz}\ \emph {et~al.}(1999)\citenamefont
  {Elowitz}, \citenamefont {Surette}, \citenamefont {Wolf}, \citenamefont
  {Stock},\ and\ \citenamefont {Leibler}}]{Elowitz1999}%
  \BibitemOpen
  \bibfield  {author} {\bibinfo {author} {\bibfnamefont {M.~B.}\ \bibnamefont
  {Elowitz}}, \bibinfo {author} {\bibfnamefont {M.~G.}\ \bibnamefont
  {Surette}}, \bibinfo {author} {\bibfnamefont {P.~E.}\ \bibnamefont {Wolf}},
  \bibinfo {author} {\bibfnamefont {J.~B.}\ \bibnamefont {Stock}}, \ and\
  \bibinfo {author} {\bibfnamefont {S.}~\bibnamefont {Leibler}},\ }\href@noop
  {} {\bibfield  {journal} {\bibinfo  {journal} {J. Bacteriol.}\ }\textbf
  {\bibinfo {volume} {181}},\ \bibinfo {pages} {197} (\bibinfo {year}
  {1999})}\BibitemShut {NoStop}%
\bibitem [{\citenamefont {Bakshi}\ \emph {et~al.}(2011)\citenamefont {Bakshi},
  \citenamefont {Bratton},\ and\ \citenamefont {Weisshaar}}]{Bakshi2011}%
  \BibitemOpen
  \bibfield  {author} {\bibinfo {author} {\bibfnamefont {S.}~\bibnamefont
  {Bakshi}}, \bibinfo {author} {\bibfnamefont {S.~P.}\ \bibnamefont {Bratton}},
  \ and\ \bibinfo {author} {\bibfnamefont {J.~C.}\ \bibnamefont {Weisshaar}},\
  }\href {\doibase 10.1016/j.bpj.2011.10.013} {\bibfield  {journal} {\bibinfo
  {journal} {Biophys. J.}\ }\textbf {\bibinfo {volume} {101}},\ \bibinfo
  {pages} {2535} (\bibinfo {year} {2011})}\BibitemShut {NoStop}%
\bibitem [{\citenamefont {English}\ \emph {et~al.}(2011)\citenamefont
  {English}, \citenamefont {Hauryliuk}, \citenamefont {Sanamrad}, \citenamefont
  {Tankov}, \citenamefont {Dekker},\ and\ \citenamefont {Elf}}]{English2011}%
  \BibitemOpen
  \bibfield  {author} {\bibinfo {author} {\bibfnamefont {B.}~\bibnamefont
  {English}}, \bibinfo {author} {\bibfnamefont {V.}~\bibnamefont {Hauryliuk}},
  \bibinfo {author} {\bibfnamefont {A.}~\bibnamefont {Sanamrad}}, \bibinfo
  {author} {\bibfnamefont {S.}~\bibnamefont {Tankov}}, \bibinfo {author}
  {\bibfnamefont {N.}~\bibnamefont {Dekker}}, \ and\ \bibinfo {author}
  {\bibfnamefont {J.}~\bibnamefont {Elf}},\ }\href {\doibase
  10.1073/pnas.1102255108} {\bibfield  {journal} {\bibinfo  {journal} {Proc.
  Natl. Acad. Sci. USA}\ }\textbf {\bibinfo {volume} {108}},\ \bibinfo {pages}
  {E365} (\bibinfo {year} {2011})}\BibitemShut {NoStop}%
\bibitem [{\citenamefont {Coquel}\ \emph {et~al.}(2013)\citenamefont {Coquel},
  \citenamefont {Jacob}, \citenamefont {Primet}, \citenamefont {Demarez},
  \citenamefont {Dimiccoli}, \citenamefont {Julou}, \citenamefont {Moisan},
  \citenamefont {Lindner},\ and\ \citenamefont {Berry}}]{Coquel2013}%
  \BibitemOpen
  \bibfield  {author} {\bibinfo {author} {\bibfnamefont {A.}~\bibnamefont
  {Coquel}}, \bibinfo {author} {\bibfnamefont {J.}~\bibnamefont {Jacob}},
  \bibinfo {author} {\bibfnamefont {M.}~\bibnamefont {Primet}}, \bibinfo
  {author} {\bibfnamefont {A.}~\bibnamefont {Demarez}}, \bibinfo {author}
  {\bibfnamefont {M.}~\bibnamefont {Dimiccoli}}, \bibinfo {author}
  {\bibfnamefont {T.}~\bibnamefont {Julou}}, \bibinfo {author} {\bibfnamefont
  {L.}~\bibnamefont {Moisan}}, \bibinfo {author} {\bibfnamefont
  {A.}~\bibnamefont {Lindner}}, \ and\ \bibinfo {author} {\bibfnamefont
  {H.}~\bibnamefont {Berry}},\ }\href {\doibase 10.1371/journal.pcbi.1003038}
  {\bibfield  {journal} {\bibinfo  {journal} {PLoS Computational Biology}\
  }\textbf {\bibinfo {volume} {9}},\ \bibinfo {pages} {e1003038} (\bibinfo
  {year} {2013})}\BibitemShut {NoStop}%
\bibitem [{\citenamefont {Weiss}\ \emph {et~al.}(2004)\citenamefont {Weiss},
  \citenamefont {Elsner}, \citenamefont {Kartberg},\ and\ \citenamefont
  {Nilsson}}]{Weiss2004}%
  \BibitemOpen
  \bibfield  {author} {\bibinfo {author} {\bibfnamefont {M.}~\bibnamefont
  {Weiss}}, \bibinfo {author} {\bibfnamefont {M.}~\bibnamefont {Elsner}},
  \bibinfo {author} {\bibfnamefont {F.}~\bibnamefont {Kartberg}}, \ and\
  \bibinfo {author} {\bibfnamefont {T.}~\bibnamefont {Nilsson}},\ }\href
  {\doibase 10.1529/biophysj.104.044263} {\bibfield  {journal} {\bibinfo
  {journal} {Biophys. J.}\ }\textbf {\bibinfo {volume} {87}},\ \bibinfo {pages}
  {3518} (\bibinfo {year} {2004})}\BibitemShut {NoStop}%
\bibitem [{\citenamefont {Guigas}\ \emph
  {et~al.}(2007{\natexlab{a}})\citenamefont {Guigas}, \citenamefont {Kalla},\
  and\ \citenamefont {Weiss}}]{Guigas2007}%
  \BibitemOpen
  \bibfield  {author} {\bibinfo {author} {\bibfnamefont {G.}~\bibnamefont
  {Guigas}}, \bibinfo {author} {\bibfnamefont {C.}~\bibnamefont {Kalla}}, \
  and\ \bibinfo {author} {\bibfnamefont {M.}~\bibnamefont {Weiss}},\ }\href
  {\doibase 10.1016/j.febslet.2007.09.054} {\bibfield  {journal} {\bibinfo
  {journal} {FEBS Lett.}\ }\textbf {\bibinfo {volume} {581}},\ \bibinfo {pages}
  {5094} (\bibinfo {year} {2007}{\natexlab{a}})}\BibitemShut {NoStop}%
\bibitem [{\citenamefont {Guigas}\ \emph
  {et~al.}(2007{\natexlab{b}})\citenamefont {Guigas}, \citenamefont {Kalla},\
  and\ \citenamefont {Weiss}}]{Guigas2007a}%
  \BibitemOpen
  \bibfield  {author} {\bibinfo {author} {\bibfnamefont {G.}~\bibnamefont
  {Guigas}}, \bibinfo {author} {\bibfnamefont {C.}~\bibnamefont {Kalla}}, \
  and\ \bibinfo {author} {\bibfnamefont {M.}~\bibnamefont {Weiss}},\ }\href
  {\doibase 10.1529/biophysj.106.099267} {\bibfield  {journal} {\bibinfo
  {journal} {Biophys J}\ }\textbf {\bibinfo {volume} {93}},\ \bibinfo {pages}
  {316} (\bibinfo {year} {2007}{\natexlab{b}})}\BibitemShut {NoStop}%
\bibitem [{\citenamefont {Murase}\ \emph {et~al.}(2004)\citenamefont {Murase},
  \citenamefont {Fujiwara}, \citenamefont {Umemura}, \citenamefont {Suzuki},
  \citenamefont {Iino}, \citenamefont {Yamashita}, \citenamefont {Saito},
  \citenamefont {Murakoshi}, \citenamefont {Ritchie},\ and\ \citenamefont
  {Kusumi}}]{Murase2004}%
  \BibitemOpen
  \bibfield  {author} {\bibinfo {author} {\bibfnamefont {K.}~\bibnamefont
  {Murase}}, \bibinfo {author} {\bibfnamefont {T.}~\bibnamefont {Fujiwara}},
  \bibinfo {author} {\bibfnamefont {Y.}~\bibnamefont {Umemura}}, \bibinfo
  {author} {\bibfnamefont {K.}~\bibnamefont {Suzuki}}, \bibinfo {author}
  {\bibfnamefont {R.}~\bibnamefont {Iino}}, \bibinfo {author} {\bibfnamefont
  {H.}~\bibnamefont {Yamashita}}, \bibinfo {author} {\bibfnamefont
  {M.}~\bibnamefont {Saito}}, \bibinfo {author} {\bibfnamefont
  {H.}~\bibnamefont {Murakoshi}}, \bibinfo {author} {\bibfnamefont
  {K.}~\bibnamefont {Ritchie}}, \ and\ \bibinfo {author} {\bibfnamefont
  {A.}~\bibnamefont {Kusumi}},\ }\href {\doibase 10.1529/biophysj.103.035717}
  {\bibfield  {journal} {\bibinfo  {journal} {Biophys J}\ }\textbf {\bibinfo
  {volume} {86}},\ \bibinfo {pages} {4075} (\bibinfo {year}
  {2004})}\BibitemShut {NoStop}%
\bibitem [{\citenamefont {Feder}\ \emph {et~al.}(1996)\citenamefont {Feder},
  \citenamefont {Brust-Mascher}, \citenamefont {Slattery}, \citenamefont
  {Baird},\ and\ \citenamefont {Webb}}]{Feder1996}%
  \BibitemOpen
  \bibfield  {author} {\bibinfo {author} {\bibfnamefont {T.~J.}\ \bibnamefont
  {Feder}}, \bibinfo {author} {\bibfnamefont {I.}~\bibnamefont
  {Brust-Mascher}}, \bibinfo {author} {\bibfnamefont {J.~P.}\ \bibnamefont
  {Slattery}}, \bibinfo {author} {\bibfnamefont {B.}~\bibnamefont {Baird}}, \
  and\ \bibinfo {author} {\bibfnamefont {W.~W.}\ \bibnamefont {Webb}},\ }\href
  {\doibase 10.1016/S0006-3495(96)79846-6} {\bibfield  {journal} {\bibinfo
  {journal} {Biophys J}\ }\textbf {\bibinfo {volume} {70}},\ \bibinfo {pages}
  {2767} (\bibinfo {year} {1996})}\BibitemShut {NoStop}%
\bibitem [{\citenamefont {Vrljic}\ \emph {et~al.}(2002)\citenamefont {Vrljic},
  \citenamefont {Nishimura}, \citenamefont {Brasselet}, \citenamefont
  {Moerner},\ and\ \citenamefont {McConnell}}]{Vrljic2002}%
  \BibitemOpen
  \bibfield  {author} {\bibinfo {author} {\bibfnamefont {M.}~\bibnamefont
  {Vrljic}}, \bibinfo {author} {\bibfnamefont {S.~Y.}\ \bibnamefont
  {Nishimura}}, \bibinfo {author} {\bibfnamefont {S.}~\bibnamefont
  {Brasselet}}, \bibinfo {author} {\bibfnamefont {W.~E.}\ \bibnamefont
  {Moerner}}, \ and\ \bibinfo {author} {\bibfnamefont {H.~M.}\ \bibnamefont
  {McConnell}},\ }\href {\doibase 10.1016/S0006-3495(02)75277-6} {\bibfield
  {journal} {\bibinfo  {journal} {Biophys. J.}\ }\textbf {\bibinfo {volume}
  {83}},\ \bibinfo {pages} {2681} (\bibinfo {year} {2002})}\BibitemShut
  {NoStop}%
\bibitem [{\citenamefont {Dix}\ and\ \citenamefont {Verkman}(2008)}]{Dix2008}%
  \BibitemOpen
  \bibfield  {author} {\bibinfo {author} {\bibfnamefont {J.~A.}\ \bibnamefont
  {Dix}}\ and\ \bibinfo {author} {\bibfnamefont {A.~S.}\ \bibnamefont
  {Verkman}},\ }\href@noop {} {\bibfield  {journal} {\bibinfo  {journal} {Annu
  Rev Biophys}\ }\textbf {\bibinfo {volume} {37}},\ \bibinfo {pages} {247}
  (\bibinfo {year} {2008})}\BibitemShut {NoStop}%
\bibitem [{\citenamefont {Condamin}\ \emph {et~al.}(2008)\citenamefont
  {Condamin}, \citenamefont {Tejedor}, \citenamefont {Voituriez}, \citenamefont
  {B\'{e}nichou},\ and\ \citenamefont {Klafter}}]{Condamin2008}%
  \BibitemOpen
  \bibfield  {author} {\bibinfo {author} {\bibfnamefont {S.}~\bibnamefont
  {Condamin}}, \bibinfo {author} {\bibfnamefont {V.}~\bibnamefont {Tejedor}},
  \bibinfo {author} {\bibfnamefont {R.}~\bibnamefont {Voituriez}}, \bibinfo
  {author} {\bibfnamefont {O.}~\bibnamefont {B\'{e}nichou}}, \ and\ \bibinfo
  {author} {\bibfnamefont {J.}~\bibnamefont {Klafter}},\ }\href@noop {}
  {\bibfield  {journal} {\bibinfo  {journal} {Proc Natl Acad Sci U S A}\
  }\textbf {\bibinfo {volume} {105}},\ \bibinfo {pages} {5675} (\bibinfo {year}
  {2008})}\BibitemShut {NoStop}%
\bibitem [{\citenamefont {Barkai}\ \emph {et~al.}(2012)\citenamefont {Barkai},
  \citenamefont {Garini},\ and\ \citenamefont
  {Metzler}}]{barkai-phystoday-2012}%
  \BibitemOpen
  \bibfield  {author} {\bibinfo {author} {\bibfnamefont {E.}~\bibnamefont
  {Barkai}}, \bibinfo {author} {\bibfnamefont {Y.}~\bibnamefont {Garini}}, \
  and\ \bibinfo {author} {\bibfnamefont {R.}~\bibnamefont {Metzler}},\ }\href
  {\doibase 10.1063/pt.3.1677} {\bibfield  {journal} {\bibinfo  {journal}
  {Physics Today}\ }\textbf {\bibinfo {volume} {65}},\ \bibinfo {pages} {29+}
  (\bibinfo {year} {2012})}\BibitemShut {NoStop}%
\bibitem [{\citenamefont {Bouchaud}\ and\ \citenamefont
  {Georges}(1990)}]{Bouchaud1990}%
  \BibitemOpen
  \bibfield  {author} {\bibinfo {author} {\bibfnamefont {J.-P.}\ \bibnamefont
  {Bouchaud}}\ and\ \bibinfo {author} {\bibfnamefont {A.}~\bibnamefont
  {Georges}},\ }\href@noop {} {\bibfield  {journal} {\bibinfo  {journal}
  {Physics Reports}\ }\textbf {\bibinfo {volume} {195}},\ \bibinfo {pages} {127
  } (\bibinfo {year} {1990})}\BibitemShut {NoStop}%
\bibitem [{\citenamefont {Metzler}\ and\ \citenamefont
  {Klafter}(2000)}]{Metzler2000}%
  \BibitemOpen
  \bibfield  {author} {\bibinfo {author} {\bibfnamefont {R.}~\bibnamefont
  {Metzler}}\ and\ \bibinfo {author} {\bibfnamefont {J.}~\bibnamefont
  {Klafter}},\ }\href {\doibase DOI: 10.1016/S0370-1573(00)00070-3} {\bibfield
  {journal} {\bibinfo  {journal} {Physics Reports}\ }\textbf {\bibinfo {volume}
  {339}},\ \bibinfo {pages} {1 } (\bibinfo {year} {2000})}\BibitemShut
  {NoStop}%
\bibitem [{\citenamefont {Renner}\ \emph {et~al.}(2005)\citenamefont {Renner},
  \citenamefont {Schutz},\ and\ \citenamefont {Vojta}}]{Renner2005}%
  \BibitemOpen
  \bibfield  {author} {\bibinfo {author} {\bibfnamefont {U.}~\bibnamefont
  {Renner}}, \bibinfo {author} {\bibfnamefont {G.~M.}\ \bibnamefont {Schutz}},
  \ and\ \bibinfo {author} {\bibfnamefont {G.}~\bibnamefont {Vojta}},\ }in\
  \href@noop {} {\emph {\bibinfo {booktitle} {Diffusion in Condensed
  Matter}}},\ \bibinfo {editor} {edited by\ \bibinfo {editor} {\bibfnamefont
  {P.}~\bibnamefont {Heitjans}}\ and\ \bibinfo {editor} {\bibfnamefont
  {J.}~\bibnamefont {Karger}}}\ (\bibinfo  {publisher} {Springer Berlin
  Heidelberg},\ \bibinfo {year} {2005})\ pp.\ \bibinfo {pages}
  {793--811}\BibitemShut {NoStop}%
\bibitem [{\citenamefont {Saxton}(1994)}]{Saxton1994}%
  \BibitemOpen
  \bibfield  {author} {\bibinfo {author} {\bibfnamefont {M.~J.}\ \bibnamefont
  {Saxton}},\ }\href@noop {} {\bibfield  {journal} {\bibinfo  {journal}
  {Biophys. J.}\ }\textbf {\bibinfo {volume} {66}},\ \bibinfo {pages} {394}
  (\bibinfo {year} {1994})}\BibitemShut {NoStop}%
\bibitem [{\citenamefont {Berry}(2002)}]{Berry2002}%
  \BibitemOpen
  \bibfield  {author} {\bibinfo {author} {\bibfnamefont {H.}~\bibnamefont
  {Berry}},\ }\href@noop {} {\bibfield  {journal} {\bibinfo  {journal} {Biophys
  J}\ }\textbf {\bibinfo {volume} {83}},\ \bibinfo {pages} {1891} (\bibinfo
  {year} {2002})}\BibitemShut {NoStop}%
\bibitem [{\citenamefont {Kammerer}\ \emph {et~al.}(2008)\citenamefont
  {Kammerer}, \citenamefont {H\"ofling},\ and\ \citenamefont
  {Franosch}}]{Kammerer2008}%
  \BibitemOpen
  \bibfield  {author} {\bibinfo {author} {\bibfnamefont {A.}~\bibnamefont
  {Kammerer}}, \bibinfo {author} {\bibfnamefont {F.}~\bibnamefont {H\"ofling}},
  \ and\ \bibinfo {author} {\bibfnamefont {T.}~\bibnamefont {Franosch}},\
  }\href@noop {} {\bibfield  {journal} {\bibinfo  {journal} {Europhys. Lett.}\
  }\textbf {\bibinfo {volume} {84}},\ \bibinfo {pages} {66002} (\bibinfo {year}
  {2008})}\BibitemShut {NoStop}%
\bibitem [{\citenamefont {H\"ofling}\ \emph {et~al.}(2008)\citenamefont
  {H\"ofling}, \citenamefont {Munk}, \citenamefont {Frey},\ and\ \citenamefont
  {Franosch}}]{Hoefling2008a}%
  \BibitemOpen
  \bibfield  {author} {\bibinfo {author} {\bibfnamefont {F.}~\bibnamefont
  {H\"ofling}}, \bibinfo {author} {\bibfnamefont {T.}~\bibnamefont {Munk}},
  \bibinfo {author} {\bibfnamefont {E.}~\bibnamefont {Frey}}, \ and\ \bibinfo
  {author} {\bibfnamefont {T.}~\bibnamefont {Franosch}},\ }\href {\doibase
  10.1063/1.2901170} {\bibfield  {journal} {\bibinfo  {journal} {J Chem Phys}\
  }\textbf {\bibinfo {volume} {128}},\ \bibinfo {pages} {164517} (\bibinfo
  {year} {2008})}\BibitemShut {NoStop}%
\bibitem [{\citenamefont {Bauer}\ \emph {et~al.}(2010)\citenamefont {Bauer},
  \citenamefont {H\"ofling}, \citenamefont {Munk}, \citenamefont {Frey},\ and\
  \citenamefont {Franosch}}]{Bauer2010}%
  \BibitemOpen
  \bibfield  {author} {\bibinfo {author} {\bibfnamefont {T.}~\bibnamefont
  {Bauer}}, \bibinfo {author} {\bibfnamefont {F.}~\bibnamefont {H\"ofling}},
  \bibinfo {author} {\bibfnamefont {T.}~\bibnamefont {Munk}}, \bibinfo {author}
  {\bibfnamefont {E.}~\bibnamefont {Frey}}, \ and\ \bibinfo {author}
  {\bibfnamefont {T.}~\bibnamefont {Franosch}},\ }\href {\doibase
  10.1140/epjst/e2010-01313-1} {\bibfield  {journal} {\bibinfo  {journal} {Eur.
  Phys. J. Special Topics}\ }\textbf {\bibinfo {volume} {189}},\ \bibinfo
  {pages} {103} (\bibinfo {year} {2010})}\BibitemShut {NoStop}%
\bibitem [{\citenamefont {Soula}\ \emph {et~al.}(2013)\citenamefont {Soula},
  \citenamefont {Car\'{e}}, \citenamefont {Beslon},\ and\ \citenamefont
  {Berry}}]{Soula2013}%
  \BibitemOpen
  \bibfield  {author} {\bibinfo {author} {\bibfnamefont {H.}~\bibnamefont
  {Soula}}, \bibinfo {author} {\bibfnamefont {B.}~\bibnamefont {Car\'{e}}},
  \bibinfo {author} {\bibfnamefont {G.}~\bibnamefont {Beslon}}, \ and\ \bibinfo
  {author} {\bibfnamefont {H.}~\bibnamefont {Berry}},\ }\href@noop {}
  {\bibfield  {journal} {\bibinfo  {journal} {Biophys J}\ } (\bibinfo {year}
  {2013})},\ \bibinfo {note} {in press}\BibitemShut {NoStop}%
\bibitem [{\citenamefont {Tremmel}\ \emph {et~al.}(2003)\citenamefont
  {Tremmel}, \citenamefont {Kirchhoff}, \citenamefont {Weis},\ and\
  \citenamefont {Farquhar}}]{Tremmel2003}%
  \BibitemOpen
  \bibfield  {author} {\bibinfo {author} {\bibfnamefont {I.~G.}\ \bibnamefont
  {Tremmel}}, \bibinfo {author} {\bibfnamefont {H.}~\bibnamefont {Kirchhoff}},
  \bibinfo {author} {\bibfnamefont {E.}~\bibnamefont {Weis}}, \ and\ \bibinfo
  {author} {\bibfnamefont {G.~D.}\ \bibnamefont {Farquhar}},\ }\href@noop {}
  {\bibfield  {journal} {\bibinfo  {journal} {Biochim Biophys Acta}\ }\textbf
  {\bibinfo {volume} {1607}},\ \bibinfo {pages} {97} (\bibinfo {year}
  {2003})}\BibitemShut {NoStop}%
\bibitem [{\citenamefont {McGuffee}\ and\ \citenamefont
  {Elcock}(2010)}]{McGuffee2010}%
  \BibitemOpen
  \bibfield  {author} {\bibinfo {author} {\bibfnamefont {S.~R.}\ \bibnamefont
  {McGuffee}}\ and\ \bibinfo {author} {\bibfnamefont {A.~H.}\ \bibnamefont
  {Elcock}},\ }\href@noop {} {\bibfield  {journal} {\bibinfo  {journal} {PLoS
  Comput Biol}\ }\textbf {\bibinfo {volume} {6}},\ \bibinfo {pages} {e1000694}
  (\bibinfo {year} {2010})}\BibitemShut {NoStop}%
\bibitem [{\citenamefont {Nenninger}\ \emph {et~al.}(2010)\citenamefont
  {Nenninger}, \citenamefont {Mastroianni},\ and\ \citenamefont
  {Mullineaux}}]{Nenninger2010}%
  \BibitemOpen
  \bibfield  {author} {\bibinfo {author} {\bibfnamefont {A.}~\bibnamefont
  {Nenninger}}, \bibinfo {author} {\bibfnamefont {G.}~\bibnamefont
  {Mastroianni}}, \ and\ \bibinfo {author} {\bibfnamefont {C.~W.}\ \bibnamefont
  {Mullineaux}},\ }\href {\doibase 10.1128/JB.00284-10} {\bibfield  {journal}
  {\bibinfo  {journal} {J Bacteriol}\ }\textbf {\bibinfo {volume} {192}},\
  \bibinfo {pages} {4535} (\bibinfo {year} {2010})}\BibitemShut {NoStop}%
\bibitem [{\citenamefont {Jacobson}\ \emph {et~al.}(1987)\citenamefont
  {Jacobson}, \citenamefont {Ishihara},\ and\ \citenamefont
  {Inman}}]{Jacobson1987}%
  \BibitemOpen
  \bibfield  {author} {\bibinfo {author} {\bibfnamefont {K.}~\bibnamefont
  {Jacobson}}, \bibinfo {author} {\bibfnamefont {A.}~\bibnamefont {Ishihara}},
  \ and\ \bibinfo {author} {\bibfnamefont {R.}~\bibnamefont {Inman}},\ }\href
  {\doibase 10.1146/annurev.ph.49.030187.001115} {\bibfield  {journal}
  {\bibinfo  {journal} {Annu Rev Physiol}\ }\textbf {\bibinfo {volume} {49}},\
  \bibinfo {pages} {163} (\bibinfo {year} {1987})}\BibitemShut {NoStop}%
\bibitem [{\citenamefont {H\"{o}fling}\ \emph {et~al.}(2006)\citenamefont
  {H\"{o}fling}, \citenamefont {Franosch},\ and\ \citenamefont
  {Frey}}]{Hofling2006}%
  \BibitemOpen
  \bibfield  {author} {\bibinfo {author} {\bibfnamefont {F.}~\bibnamefont
  {H\"{o}fling}}, \bibinfo {author} {\bibfnamefont {T.}~\bibnamefont
  {Franosch}}, \ and\ \bibinfo {author} {\bibfnamefont {E.}~\bibnamefont
  {Frey}},\ }\href@noop {} {\bibfield  {journal} {\bibinfo  {journal} {Physical
  Review Letters}\ }\textbf {\bibinfo {volume} {96}},\ \bibinfo {pages}
  {165901} (\bibinfo {year} {2006})}\BibitemShut {NoStop}%
\bibitem [{\citenamefont {Gillespie}(1996)}]{Gillespie1996}%
  \BibitemOpen
  \bibfield  {author} {\bibinfo {author} {\bibfnamefont {D.~T.}\ \bibnamefont
  {Gillespie}},\ }\href@noop {} {\bibfield  {journal} {\bibinfo  {journal}
  {Phys. Rev. E}\ }\textbf {\bibinfo {volume} {54}},\ \bibinfo {pages} {2084}
  (\bibinfo {year} {1996})}\BibitemShut {NoStop}%
\bibitem [{\citenamefont {Saxton}(1990)}]{Saxton1990}%
  \BibitemOpen
  \bibfield  {author} {\bibinfo {author} {\bibfnamefont {M.}~\bibnamefont
  {Saxton}},\ }\href@noop {} {\bibfield  {journal} {\bibinfo  {journal}
  {Biophys. J.}\ }\textbf {\bibinfo {volume} {58}},\ \bibinfo {pages} {1303}
  (\bibinfo {year} {1990})}\BibitemShut {NoStop}%
\bibitem [{\citenamefont {Scheuring}\ and\ \citenamefont
  {Sturgis}(2006)}]{Scheuring2006}%
  \BibitemOpen
  \bibfield  {author} {\bibinfo {author} {\bibfnamefont {S.}~\bibnamefont
  {Scheuring}}\ and\ \bibinfo {author} {\bibfnamefont {J.~N.}\ \bibnamefont
  {Sturgis}},\ }\href {\doibase 10.1529/biophysj.106.083709} {\bibfield
  {journal} {\bibinfo  {journal} {Biophys J}\ }\textbf {\bibinfo {volume}
  {91}},\ \bibinfo {pages} {3707} (\bibinfo {year} {2006})}\BibitemShut
  {NoStop}%
\bibitem [{\citenamefont {Ehrig}\ \emph {et~al.}(2011)\citenamefont {Ehrig},
  \citenamefont {Petrov},\ and\ \citenamefont {Schwille}}]{Ehrig2011}%
  \BibitemOpen
  \bibfield  {author} {\bibinfo {author} {\bibfnamefont {J.}~\bibnamefont
  {Ehrig}}, \bibinfo {author} {\bibfnamefont {E.~P.}\ \bibnamefont {Petrov}}, \
  and\ \bibinfo {author} {\bibfnamefont {P.}~\bibnamefont {Schwille}},\ }\href
  {\doibase 10.1016/j.bpj.2010.11.002} {\bibfield  {journal} {\bibinfo
  {journal} {Biophys J}\ }\textbf {\bibinfo {volume} {100}},\ \bibinfo {pages}
  {80} (\bibinfo {year} {2011})}\BibitemShut {NoStop}%
\bibitem [{\citenamefont {K\"{u}hn}\ \emph {et~al.}(2011)\citenamefont
  {K\"{u}hn}, \citenamefont {Ihalainen}, \citenamefont {Hyv\"{a}luoma},
  \citenamefont {Dross}, \citenamefont {Willman}, \citenamefont {Langowski},
  \citenamefont {Vihinen-Ranta},\ and\ \citenamefont {Timonen}}]{Kuhn2011}%
  \BibitemOpen
  \bibfield  {author} {\bibinfo {author} {\bibfnamefont {T.}~\bibnamefont
  {K\"{u}hn}}, \bibinfo {author} {\bibfnamefont {T.~O.}\ \bibnamefont
  {Ihalainen}}, \bibinfo {author} {\bibfnamefont {J.}~\bibnamefont
  {Hyv\"{a}luoma}}, \bibinfo {author} {\bibfnamefont {N.}~\bibnamefont
  {Dross}}, \bibinfo {author} {\bibfnamefont {S.~F.}\ \bibnamefont {Willman}},
  \bibinfo {author} {\bibfnamefont {J.}~\bibnamefont {Langowski}}, \bibinfo
  {author} {\bibfnamefont {M.}~\bibnamefont {Vihinen-Ranta}}, \ and\ \bibinfo
  {author} {\bibfnamefont {J.}~\bibnamefont {Timonen}},\ }\href {\doibase
  10.1371/journal.pone.0022962} {\bibfield  {journal} {\bibinfo  {journal}
  {PLoS ONE}\ }\textbf {\bibinfo {volume} {6}},\ \bibinfo {pages} {e22962}
  (\bibinfo {year} {2011})}\BibitemShut {NoStop}%
\bibitem [{\citenamefont {de~Gennes}(1983)}]{Gennes1983}%
  \BibitemOpen
  \bibfield  {author} {\bibinfo {author} {\bibfnamefont {P.}~\bibnamefont
  {de~Gennes}},\ }\href@noop {} {\bibfield  {journal} {\bibinfo  {journal}
  {C.R.Acad. Sc. Paris SerII}\ }\textbf {\bibinfo {volume} {296}},\ \bibinfo
  {pages} {881} (\bibinfo {year} {1983})}\BibitemShut {NoStop}%
\bibitem [{\citenamefont {Condamin}\ \emph {et~al.}(2007)\citenamefont
  {Condamin}, \citenamefont {Bénichou}, \citenamefont {Tejedor}, \citenamefont
  {Voituriez},\ and\ \citenamefont {Klafter}}]{Condamin2007}%
  \BibitemOpen
  \bibfield  {author} {\bibinfo {author} {\bibfnamefont {S.}~\bibnamefont
  {Condamin}}, \bibinfo {author} {\bibfnamefont {O.}~\bibnamefont {Bénichou}},
  \bibinfo {author} {\bibfnamefont {V.}~\bibnamefont {Tejedor}}, \bibinfo
  {author} {\bibfnamefont {R.}~\bibnamefont {Voituriez}}, \ and\ \bibinfo
  {author} {\bibfnamefont {J.}~\bibnamefont {Klafter}},\ }\href {\doibase
  10.1038/nature06201} {\bibfield  {journal} {\bibinfo  {journal} {Nature}\
  }\textbf {\bibinfo {volume} {450}},\ \bibinfo {pages} {77} (\bibinfo {year}
  {2007})}\BibitemShut {NoStop}%
\end{thebibliography}%

\end{document}